\documentclass[twocolumn, astrosymb, twocolappendix]{aastex7}

\usepackage{graphicx}
\usepackage{subcaption}
\usepackage{placeins}
\usepackage{enumerate}
\usepackage{enumitem}
\usepackage{mathtools}
\usepackage{upgreek}
\usepackage{gensymb}
\usepackage{multirow}
\usepackage{makecell}
\usepackage{amssymb}
\usepackage[T1]{fontenc}
\usepackage{amsfonts}
\usepackage[version=4]{mhchem}
\usepackage{stmaryrd}
\usepackage{amsmath}
\usepackage{mathrsfs}
\usepackage{dsfont}
\usepackage{accents}
\graphicspath{ {./images/} }
\usepackage{hyperref}
\usepackage{colortbl}
\usepackage{array}
\usepackage{float}
\usepackage[inkscapelatex=false]{svg}

\defcitealias{drimmel2023gaia}{Gaia Collaboration 2023a}
\defcitealias{recio2023gaia}{b}
\defcitealias{schultheis2023gaia}{c}

\begin{document}

\title{Identification of Outer Galaxy Cluster Members Using Gaia DR3 and Multidimensional Simulation}

\author[0009-0009-2207-9999]{Vishwas Patel}
\altaffiliation{Current affiliation: Space Applications Centre, Ahmedabad, Indian Space Research Organization, India - 380015.}
\affiliation{Indian Institute of Space science and Technology (IIST), Thiruvananthapuram, Kerala, India - 695547}
\email[show]{vrhpatel2000@gmail.com}

\author[0000-0002-5599-4650]{Joseph L. Hora}
\affiliation{Center for Astrophysics $|$ Harvard \& Smithsonian, 60 Garden St., Cambridge, MA 01238, USA}
\email[show]{jhora@cfa.harvard.edu}

\author[0000-0002-3993-0745]{Matthew L.N. Ashby}
\affiliation{Center for Astrophysics $|$ Harvard \& Smithsonian, 60 Garden St., Cambridge, MA 01238, USA}
\email{mashby@cfa.harvard.edu}

\author[0000-0002-3477-6021]{Sarita Vig}
\affiliation{Indian Institute of Space science and Technology (IIST), Thiruvananthapuram, Kerala, India - 695547}
\email{sarita@iist.ac.in}

\correspondingauthor{Joseph L. Hora, Vishwas Patel}

\begin{abstract}

The outer Galaxy presents a distinctive environment for investigating star formation. This study develops a novel approach to identify true cluster members based on unsupervised clustering using astrometry with significant uncertainties. As a proof of concept, we analyze three outer Galactic Young Stellar Object (YSO) clusters at different distances and densities within $65^\circ < l < 265^\circ$, each known to contain $>100$ members based on the Star Formation in Outer Galaxy \citep[SFOG;][]{Winston2020} catalog. The 618 YSO clusters in the SFOG dataset were based on 2-dimensional clustering.  In this contribution, we apply the $\text{HDBSCAN}^{*}$ algorithm to the precise Gaia DR3 astrometry to assign YSO cluster membership. Monte Carlo simulation, coupled with $\text{HDBSCAN}^{*}$ (HDBSCAN-MC algorithm), addresses YSO astrometric uncertainties while 5-dimensional clustering. We introduce the Generation Of cLuster anD FIeld STar (GOLDFIST) simulation, to enable robust membership determination, performing an unsupervised clustering analysis in higher-dimensional feature space while accommodating measurement errors. In this study, we extended our approach to distant outer galaxy YSOs and clusters with larger astrometric uncertainties. The results include the discovery of new members in the previously identified clusters. We also analyze the known stars in the clusters and confirm their membership. The derived membership probabilities are included in the provided cluster catalogs. The more accurately predicted simulation distance estimates closely agree, within uncertainty limits, with the median distance estimates derived from Gaia data, and are compared with the kinematic distances from the WISE \ion{H}{2} survey. 
\end{abstract}

\keywords{\uat{Clustering}{1908} --- \uat{Young stellar objects}{1834} --- \uat{Young star clusters}{1833} --- \uat{Star forming regions}{1565} --- \uat{Computational astronomy}{293} --- \uat{Astrometry}{80}}

\section{Introduction}\label{sec:introduction}

Star clusters primarily form within cold, dense molecular clouds where gravitational collapse initiates the formation of new stars \citep{mckee2007theory}. In the initial phases, the clusters are deeply ensconced within the parental clouds and are referred to as embedded clusters \citep{lada2003embedded}.

The giant molecular clouds that form these clusters are predominantly located within the spiral arms of galaxies. The arm-to-interarm ratio of the surface molecular density can vary significantly, ranging from a few times in the inner regions of some galaxies \citep{Vogel1988, brouillet1997identification} to more than 20 times in the outer regions of the Milky Way  \citep{Digel1996}. 

Research on star formation has largely been focused on nearby clouds in the inner Galaxy, often overlooking the outer Galaxy, where physical conditions differ from the inner Galaxy \citep{heyer2015molecular}. The Milky Way's metallicity decreases with increasing Galactocentric radius \citep{rudolph1997far}. Additionally, molecular clouds in the outer Galaxy show lower temperatures \citep{mead1988molecular}, reduced cosmic-ray flux \citep{bloemen1984radial}, and lower densities. These factors lead to fewer interactions and spiral arm crossings, reducing star formation efficiency \citep{Winston2020}. An accurate assessment of their membership would be very useful in making these outer-Galaxy clusters accessible for astrophysical studies, enabling the estimation of key parameters like distance, age, and metallicity, which are vital for understanding the Galaxy’s spiral structure, star formation history, and chemical evolution. However, very few studies have focused on analyzing the membership and properties of clusters in these distant regions to date.

Traditionally, the discovery of Galactic clusters and their members relied on two-dimensional positional data and photometric information \citep{baume2003photometric, trumpler1979preliminary}. The ESA Gaia mission \citep{Perryman2001, Creevey2023}, launched in 2013, revolutionized this field. An important contribution of Gaia to Milky Way cluster studies concerns the cluster census itself. Using DR2 data, \citet{Cantat2018} could identify only 1229 of the 3000+ known clusters. Further, this study revealed that many so-called clusters were asterisms without kinematic association. \citet{Perren2020} employed the Automated Stellar Cluster Analysis pipeline (ASteCA, \citealt{Perren2015}) to simultaneously estimate parameters such as distance, reddening, total mass, age, and metallicity for 16 clusters. Gaia’s precise proper motions allow detection of sparse yet kinematically coherent star groups. While most studies focus on Gaia observables, some have explored transformed quantities like action-angle space \citep[e.g.,][]{Coronado2020,Coronado2022}.

Modern star cluster searches have greatly benefited from advanced unsupervised clustering algorithms, leading to numerous discoveries. Using a nearest-neighbor approach, \citet{He2021} identified 74 clusters, while \citet{Liu2019} proposed 76 new cluster candidates with a friend-of-friend cluster finder method. Visual inspection of proper motion diagrams by \citet{Sim2019} revealed 207 new open star clusters within 1 kpc. Recent studies by \citet{Rimoldini2023} and \citet{marton2023gaia} have employed supervised classification techniques to categorize variable sources in Gaia DR3, identifying potential Young Stellar Object (YSO) candidates among various other types of variables. In the quest for distant objects, \citet{Ferreira2020} detected 25 sources using spatial and photometric filters. Density-based algorithms like DBSCAN \citep[e.g.][]{Hunt2023, Winston2020} and the newer $\text{HDBSCAN}^{*}$ \citep{Hunt2023, Kerr2023, Qin2023} have also significantly advanced cluster detection. $\text{HDBSCAN}^{*}$ identifies natural clusters with variable densities by using minimum cluster size as a parameter, eliminating the need for DBSCAN's fixed $\epsilon$ distance threshold \citep{ester1996density}. This study utilizes the $\text{HDBSCAN}^{*}$ algorithm \citep{McInnes2017,mcinnes2017accelerated} for young cluster identification, capitalizing on its vastly superior performance and speed.

Even though Gaia DR3 provides good astrometric precision for nearby objects, there is an increase in uncertainty for distant sources, leading most studies to focus on nearby and inner Galaxy clusters. In this study, we develop and demonstrate a novel methodology to analyze distant outer Galactic clusters amidst significant astrometric uncertainties and ascertain their stellar membership. Focusing primarily on YSO clusters, we combine complementary catalogs, including Gaia DR3 (optical: \citealt{Rimoldini2023, DR3data}) and the Star Formation in the Outer Galaxy (SFOG, IR catalog: \citealt{Winston2020}) catalog for the determination of YSO memberships. 

This paper is organized as follows. Section 2 describes the data compilation from the Gaia DR3 and SFOG catalogs and the preprocessing steps for further cluster analysis. Section 3 describes the development of HDBSCAN-MC, a generalized unsupervised clustering algorithm for n-dimensional clustering considering the uncertainties. The developed Generation Of cLuster anD FIeld STar (GOLDFIST) simulation for robust cluster member detection is discussed in Section 4. Section 5 illustrates the proof of concept of the overall methodology on three outer Galactic clusters\footnote{The ``outer Galaxy" refers to the region beyond Galactocentric radius of 8 kpc, spanning approximately $65\degree < l < 265\degree$ \citep{Winston2020}. The analyzed regions include Cluster 123, the W4/W5 complex (Cluster 257), and Cluster 163, where number-based identification is according to \citet{Winston2020}.} of varied distances and densities. Finally, the conclusions are presented in Section 6. 

\section{Data Compilation and Preprocessing}\label{sec:data_compilation} 

The cross-matching and processing of data obtained from the SFOG (infrared: Spitzer—\citealt{spitzer}, WISE—\citealt{wise}, and 2MASS—\citealt{2mass}) and the Gaia DR3 catalog (optical: \citealt{gaia_2016b,gaia_2023j,gaia_2023_validation}) were conducted to analyze samples of YSOs and their clusters in the outer Galaxy.

\subsection{Catalog Cross-match with Gaia (E)DR3}\label{subsec: crossmatch_DR3}

Utilizing 2D clustering with DBSCAN, 618 YSO clusters (each with more than 5 members) were identified within the SFOG data \citep{Winston2020}, making it a valuable dataset for further investigation. Of the 47,405 YSOs listed in SFOG, we removed 142 YSOs that were duplicated and obtained a refined catalog with 47,263 unique YSO entries. Cross-matching these SFOG YSOs with the Gaia DR3 presented a challenge due to the difference in their epochs of observations. We employed a two-step approach to mitigate this issue. Initially, the SFOG catalog (J2000 epoch) was cross-matched with the Gaia EDR3 \citep[also J2000]{EDR3data}, within a 0\farcs5 search radius, and subsequently joined Gaia DR3 columns using the unique \texttt{source\_id}. The Gaia DR3 magnitude uncertainties were included from the \textit{“I/355/gaiadr3”} catalog in Vizier, where CDS provided the necessary error values.

We identified 242 "one-to-many" matches (one SFOG source matched to multiple Gaia sources) and 98 "many-to-one" matches (multiple SFOG sources matched to the same Gaia object within 0\farcs5). Among the many-to-one cases, 79 corresponded to the same YSO independently identified by the GLIMPSE and SMOG surveys, 18 by SMOG and ALLWISE, and one by SMOG and the IRAC-WISE catalog \citep{Winston2020}. SMOG data, offering broader spectral coverage, was retained, reducing the final count to 26,570 YSOs (26,669 - 98).

The primary concern for one-to-one chance alignment is the erroneous association of a foreground/background optical source with an IR source. We used photometery to resolve any such many-to-one or one-to-one chance alignment. For instance, a correct SFOG YSO-Gaia match exhibits an infrared excess, as shown in Figure \ref{fig: mag_issue_chance_good}(a) in Appendix \ref{app: sfog-gaia}. Incorrect matches were identified by comparing optical and infrared magnitudes, with criteria given by Equation 1 where $\sigma_i$ and $\sigma_j$ represent the magnitude uncertainties of the respective optical and infrared bands. Optical magnitudes refer to Gaia bands ($G_{BP}$, $G$, $G_{RP}$), while all other bands up to MIPS 24 $\mu$m are referred to as 'infrared bands'.

\begin{equation}
    \min(\text{optical mag} - 1\sigma_i) \leq \max(\text{infrared mag} + 1\sigma_j)
    \label{eq: mag_criteria}
\end{equation}

Using the criteria outlined in Equation \ref{eq: mag_criteria}, we identified a total of 94 spurious matches. Two such cases out of 94 are depicted in Figure \ref{fig: mag_issue_chance_good}(b) in Appendix \ref{app: sfog-gaia}. Notably, all of the 94 chance alignments were one-to-one matches. Hence best match candidates based on angular separation were chosen for the one-to-many cases. The final catalog was obtained by eliminating duplicated YSOs, chance alignments, and multiple matches. The refined catalog now comprises 26,668 - 246 - 98 - 94 = 26,230 SFOG YSOs with Gaia counterparts. These will be henceforth called the Gaia-matched SFOG YSOs.

\subsection{Adding Gaia-identified variable young stellar object candidates}

Gaia DR3 photometry facilitated the identification of various types of variables, including variable YSOs \citep{Rimoldini2023}.
We combined 79,375 Gaia-identified YSOs, as reported by \citet{marton2023gaia}, with SFOG YSOs to expand our study sample. This resulted in a total of 104,600 YSOs, with addition of 78,370 unique Gaia-only YSOs. We obtained Gaia DR3 astrometry and photometry for archival Gaia YSOs and updated their astrometry to the J2000 epoch using a 0\farcs05 cross-match in CDS.

\subsection{Distance estimates for Gaia identified YSOs}
 
Distance estimates for the Gaia-detected sources include the General Stellar Parameterizer from Photometry based distances (GSP Phot distance; \citealt{Andrae2023}) which are obtained from the Gaia DR3 Apsis pipeline. The geometric and photogeometric distances from Gaia data are derived by \citet{bailer2021estimating}. A comparison of these distance estimates is given in \citet{fouesneau2023gaia}. GSP Phot distances are mostly reliable within 2\;kpc, and in few cases up to 10\;kpc when the parallax over error ($\frac{\varpi}{\sigma_{\varpi}}$) is greater than 10. Geometric and photogeometric estimates perform better than the GSP Phot estimate, which is found to systematically underestimate the distances, e.g., \citetalias{drimmel2023gaia},\citetalias{recio2023gaia},\citetalias{schultheis2023gaia}. 

Out of photogeometric and geometric distances photogeometric is found to be a better estimate as analyzed in \citet{bailer2021estimating}. Therefore, we used GSP-Phot distances and uncertainties when $\frac{\varpi}{\sigma_{\varpi}} > 20$ and photogeometric distances were unavailable. Otherwise, we used geometric distances and their uncertainties. In all other cases, we preferred photogeometric distances and uncertainties whenever available.

The finalized distances were converted to the Galactocentric frame (refer Appendix \ref{app: sfog-gaia_galcen}) using the Astropy library's Galactocentric class \citep{astropy:2013}. Distance estimates for 2140 out of 104600 YSOs were not available. Hence, the Galactocentric X, Y, and Z YSO coordinates, derived from ICRS RA, Dec (J2000 epoch), and distances (pc), were appended to an updated catalog of 102,460 YSOs. Sources with renormalized unit weight errors (RUWE) $\geq 1.4$, indicating potentially problematic solutions \citep{Fabricius2021}, were excluded, yielding a refined sample of 79,654 YSO candidates. The outer Galaxy survey field is defined using the following cut \citep{Winston2020}: $(64.524179\degree<l<265.690015\degree) \ \text{and} \ (-3.179873\degree<b<3.491735\degree)$. The final Gaia-SFOG dataset used for further analysis, therefore comprises 30,738 young stellar objects.

\section{HDBSCAN-MC: Unsupervised phase space clustering with uncertainties} \label{sec:HDBSCAN_MC}

In this section, we present the HDBSCAN-MC algorithm, which we developed to assign cluster membership through unsupervised clustering in the presence of significant data uncertainties. The method can be broadly extended to n-dimensional clustering while effectively managing inherent uncertainties.

$\text{HDBSCAN}^{*}$ (Hierarchical Density-Based Spatial Clustering of Applications with Noise), developed by \citet{Campello2013}, is a sophisticated and efficient algorithm for high-dimensional clustering, demonstrating exceptional performance with Gaia data \citep{Qin2023}. $\text{HDBSCAN}^{*}$ offers intuitive parameter selection, mainly requiring the \texttt{min\_cluster} parameter, and demonstrates efficiency in the identification of sparse clusters. The primary motivation for integrating Monte Carlo simulation with $\text{HDBSCAN}^{*}$ clustering (jointly named HDBSCAN-MC) stemmed from the presence of significant uncertainty in Gaia astrometry for distant sources. The astrometric dimensions used for clustering include (X, Y, Z, $Vel_{RA}$, $Vel_{Dec}$), where X, Y, and Z denote YSO Galactocentric coordinates. The projected velocities in Right ascension ($Vel_{RA}$) and Declination ($Vel_{Dec}$) are derived by multiplying the sampled distances (pc) with the tangent of sampled proper motions in RA ($\mu_{\alpha} \cos\delta$) and Dec ($\mu_{\delta}$), respectively. 

\subsection{Integration of 3D Monte-carlo Simulations to $\text{HDBSCAN}^{*}$}\label{subsec:3d_montecarlo}

\begin{figure}[h]
    \centering
    \includegraphics[width=\linewidth]{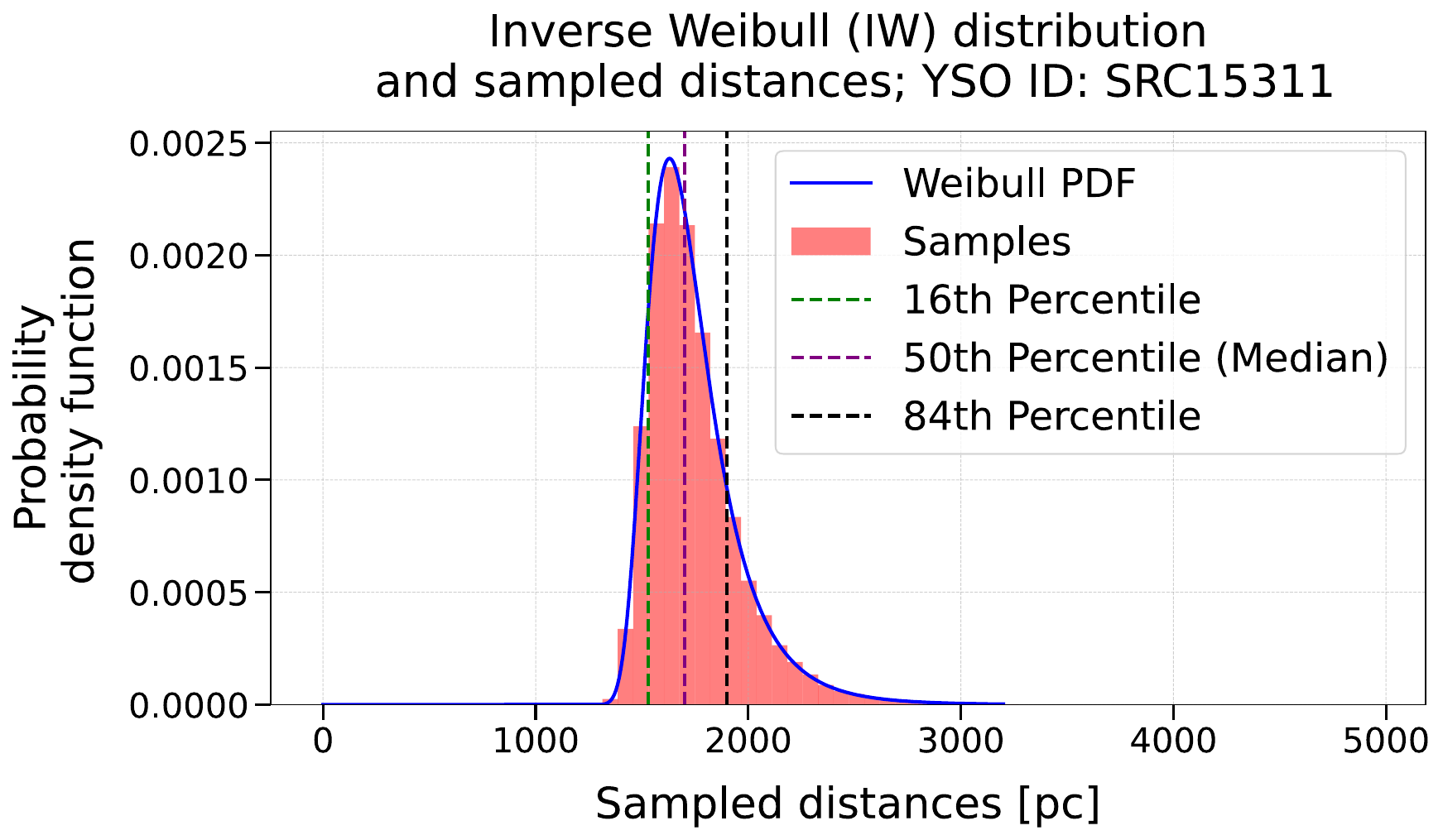} 
    \caption{Distance sampling from the Inverse Weibull distribution for Monte Carlo simulation. The blue line represents the inverse Weibull distribution derived from the $16^{th}$, $50^{th}$, and $84^{th}$ distance percentiles, indicated by the green, purple, and black dashed lines, respectively. These distance percentiles correspond to the source SRC15311. The faded red histogram illustrates $10^{5}$ sampled distances, showing that they closely follow the parent distribution.}\label{fig: dist_sampling_1}
\end{figure}

HDBSCAN-MC performs 5-dimensional $\text{HDBSCAN}^{*}$ clustering on $10^5$ mock YSO catalogs for the sky region under analysis. These mock catalogs are generated by sampling ICRS distances from an Inverse Weibull distance error distribution and proper motions in RA and Dec from their respective Gaussian error distributions, resulting in a 3D Monte Carlo sampling process. The Inverse Weibull distribution was found to best model the posterior distance distributions for both geometric and photogeometric estimates from \citet{bailer2021estimating}, particularly capturing their long-tailed nature at larger distances and skewness. Figure \ref{fig: dist_sampling_1} illustrates the Inverse Weibull distribution and demonstrates how the randomly sampled data strictly follows this distribution.

The probability density function (PDF) of the Inverse Weibull distribution is described in \citet{sherina2014generalized}. Following the same notation, the parameters $\alpha$ and $\beta$ represent the scale and shape parameters, respectively. The $16^{th}$, $50^{th}$, and $84^{th}$ distance percentiles, denoted by $d_{16}$, $d_{50}$, and $d_{84}$ respectively are the only known information about the \citet{bailer2021estimating} distance distributions. The parameters $\alpha$ and $\beta$ can be obtained by using these distance percentiles along with the distribution's cumulative distribution function (CDF) as given \citealt{sherina2014generalized}. The Equations \ref{eq: alpha} and \ref{eq: beta} are used for calculating $\alpha$ and $\beta$.

\begin{equation}
    \alpha = \frac{1}{d_{50}.log(2)^{\frac{1}{k}}}\label{eq: alpha}
\end{equation}

\begin{equation}
\beta = \frac{\log\left[\frac{\log(0.84)}{\log(0.16)}\right]}{\log\left(\frac{d_{16}}{d_{84}}\right)}\label{eq: beta}
\end{equation} 

Let $\mu$ and $\sigma$ denote the mean and standard deviation of the proper motion in the right ascension ($\rm\mu_{\alpha}cos\delta$) distribution, respectively. The values for $\mu$ and $\sigma$ for each YSO are obtained from Gaia data. Consequently, the PDF from which $\rm\mu_{\alpha}cos\delta$ values are sampled is given by Equation \ref{eq: pm_pdf}. A similar procedure is applied to sample the proper motion in declination ($\mu_{\delta}$). 

\begin{equation}
    f(x;\mu,\sigma) =\frac{1}{\sigma \sqrt{2\pi}}e^{-\frac{(x-\mu)^2}{2\sigma^2}}\label{eq: pm_pdf}
\end{equation}

If multiple clusters are detected by $\text{HDBSCAN}^{*}$ during clustering of a particular mock catalog, the largest (primary) cluster is considered for membership analysis. Although the developed code allows analysis of secondary or say tertiary clusters, none were found for the young clusters analyzed in this study. We adopt a soft clustering probability threshold of 0.1 for $\text{HDBSCAN}^{*}$, providing a balanced criterion for including or excluding peripheral cluster members.

\begin{figure*}[ht]
    \centering
    \begin{subfigure}[h]{0.49\linewidth}
        \centering
        \includegraphics[width=\linewidth]{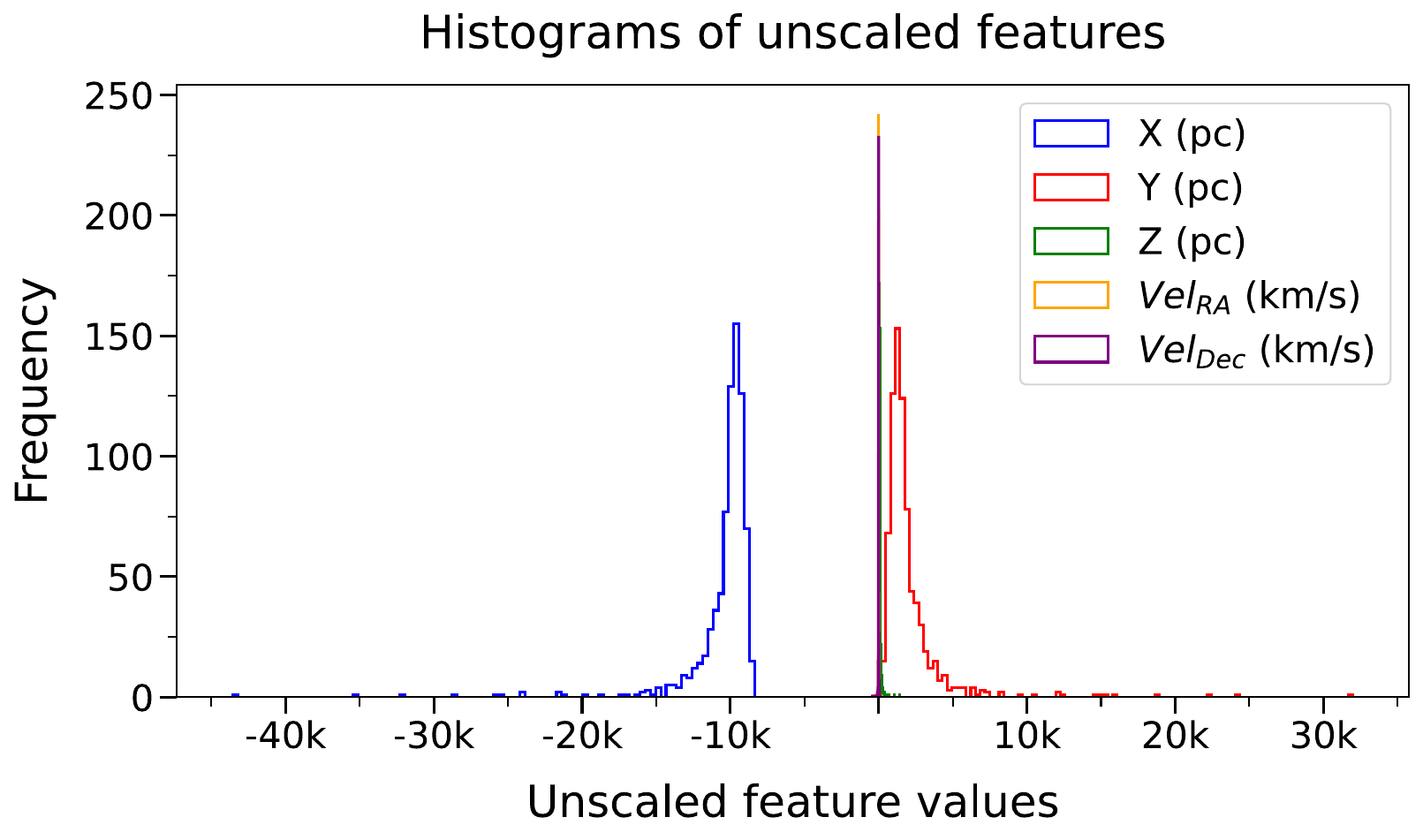}
        \caption{\centering Feature distribution before scaling.}\label{subfig: unscaled_257}
    \end{subfigure}
   \hfill
    \begin{subfigure}[h]{0.49\linewidth}
        \centering
        \includegraphics[width=\linewidth]{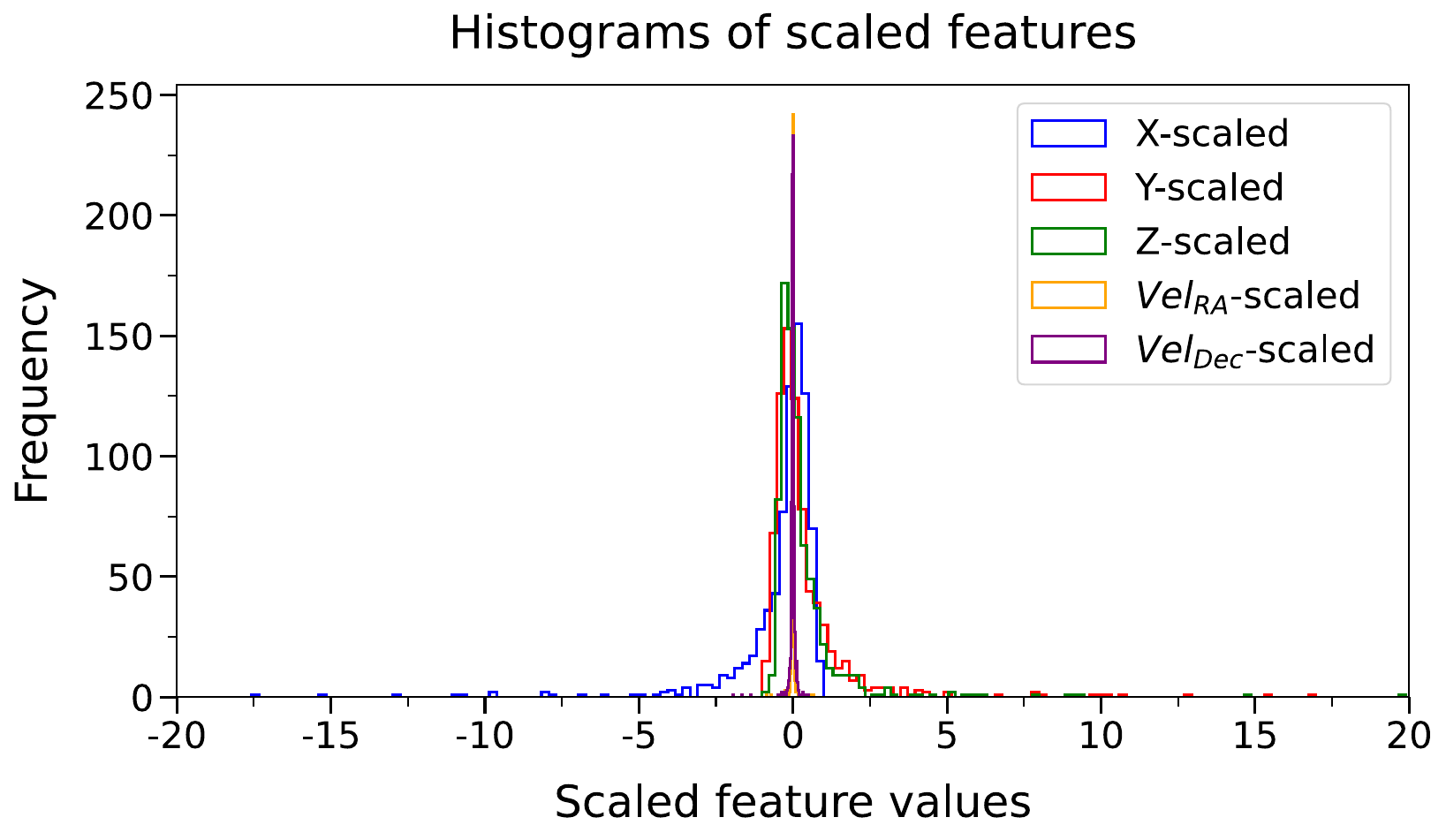}     \caption{\centering Feature distribution after scaling.}\label{subfig: scaled_257}
    \end{subfigure}
    \caption{Comparison of feature distribution for the W4/W5 complex (Cluster 257) for a mock dataset generated during the execution of HDBSCAN-MC: (a) before scaling and (b) after scaling. The X, Y, and Z coordinates (in pc) represent the YSO's Galactocentric positions derived from sampled ICRS distances, while $Vel_{RA}$ and $Vel_{Dec}$ (in km/s) represent velocities obtained from sampled ICRS YSO distances and sampled proper motions. The application of feature scaling according to Equations \ref{eq: X_scaling} and \ref{eq: vel_scaling} effectively brings the distribution of YSO astrometric features into comparable ranges, facilitating clustering with $\text{HDBSCAN}^{*}$.}\label{fig: scaled_data}
\end{figure*}

\subsection{Feature Engineering\label{subsec:feature_engineering}}

Feature engineering is a technique in machine learning that ensures all features are treated on nearly equal footing.  In the present 5-dimensional clustering the features include (X, Y, Z, $Vel_{RA}$, $Vel_{Dec}$). In distant clusters, cluster members may merge with surrounding sources in the proper motion vector plot, making proper motions and thus velocities less reliable compared to spatial dimensions. Nevertheless, it is also crucial to avoid completely isolated YSOs in the proper motion plot to be identified as cluster members. Taking into consideration the above requirements, positional dimensions (X, Y, Z) and velocity dimensions ($Vel_{RA}$, $Vel_{Dec}$) are scaled using Equations \ref{eq: X_scaling} and \ref{eq: vel_scaling}, respectively. Scaling utilized the interquartile range between the $k^{th}$ to $l^{th}$ percentiles ($IQR_{kl}$) with $k=0$, $l=50$ for position scaling, and $k=0$, $l=99.5$ for velocity scaling.

IQR represents the order of the feature, thus normalizing the median subtracted feature with IQR brings their distributions to nearly the same range, without getting affected by outliers. The proper motion distribution of YSOs in the cluster approximates a Gaussian and since the absolute value is taken, a factor of `2' is multiplied in the denominator.

\begin{equation}
    X' = \frac{X-median(X)}{IQR_{kl}(|X|)}\label{eq: X_scaling}
\end{equation}

\begin{equation}
    Vel' = \frac{Vel- median(Vel)}{2 \times IQR_{kl}(|Vel|)}\label{eq: vel_scaling}
\end{equation}

The positional dimensions, unlike velocity, were sampled from heavy-tailed Inverse Weibull distributions, extending towards larger distances. Consequently, many of the YSO distance coordinates might exhibit large negative values, particularly along the X-axis, given the consideration of the outer Galaxy. Calculating the $0^{th}$ percentile in such cases would yield the maximum distance rather than the desired least distance from the Galactic center. Therefore, to ensure an accurate representation, the absolute value was taken for the positional dimensions (X, Y, and Z) before computing the interquartile range (IQR). Figure \ref{fig: scaled_data} illustrates the scaled feature distribution for a generated mock YSO data and its comparison with unscaled features.

\begin{figure*}[!htpb]
		\centering
	\includegraphics[width=0.85\linewidth]{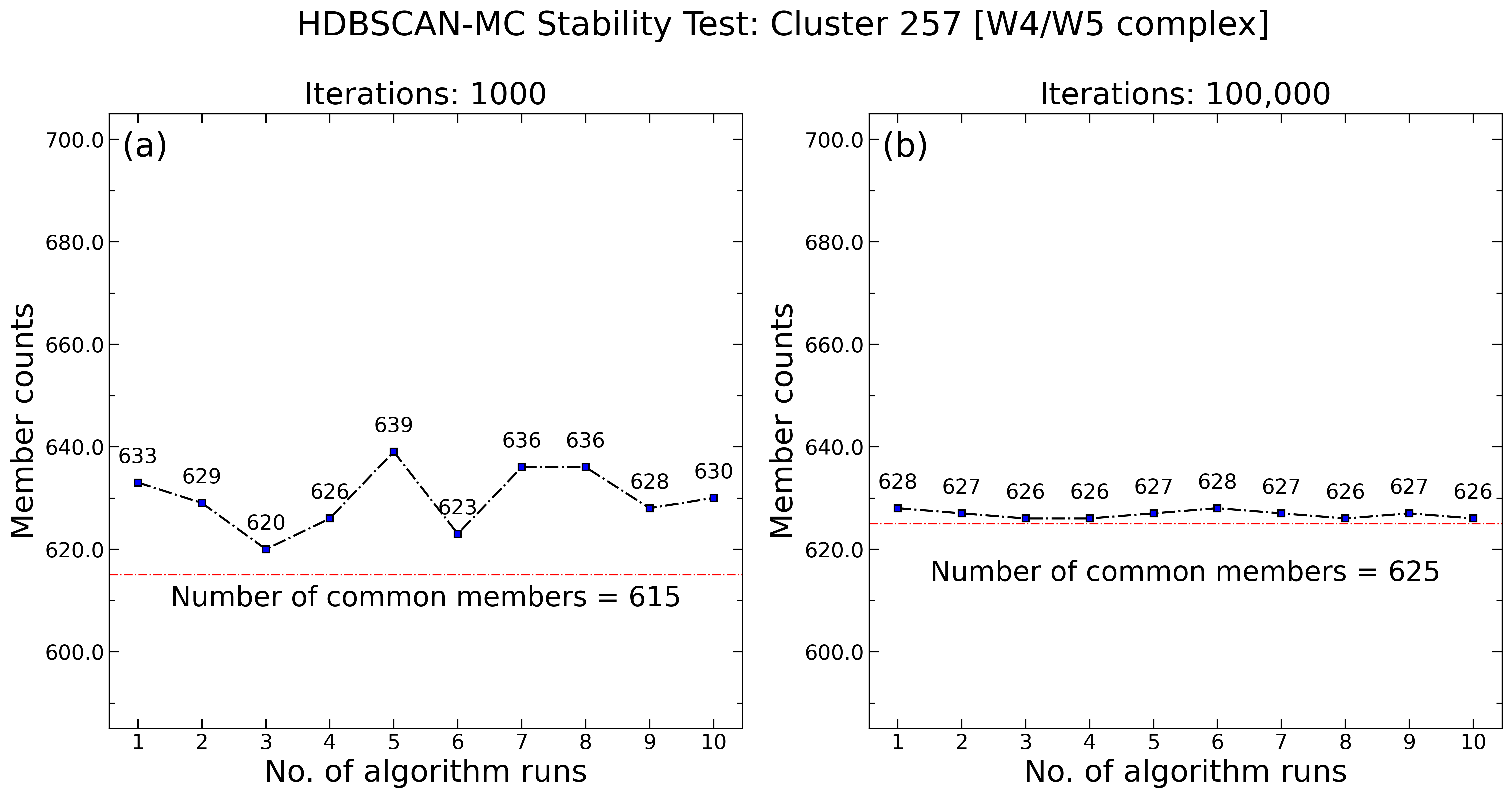} 
		\caption{HDBSCAN-MC stability test for Cluster 257 (Cluster in W4/W5 complex) (a) For 1000 iterations (b) For 100,000 iterations. The horizontal axis denotes the run for which the algorithm is repeated and the vertical axis gives the number of cluster members identified by HDBSCAN-MC in that run. The Monte Carlo threshold derived from the cluster simulation is set at 29,000 for this cluster. The dotted red line marks the number of common members across all HDBSCAN-MC runs, representing the final cluster members. The plots demonstrate the convergence of common members to a stable value as the number of Monte Carlo iterations increases.}\label{fig: st_test_257}
\end{figure*}

\begin{figure}[htpb!]
		\centering
    \includegraphics[width=0.95\linewidth]{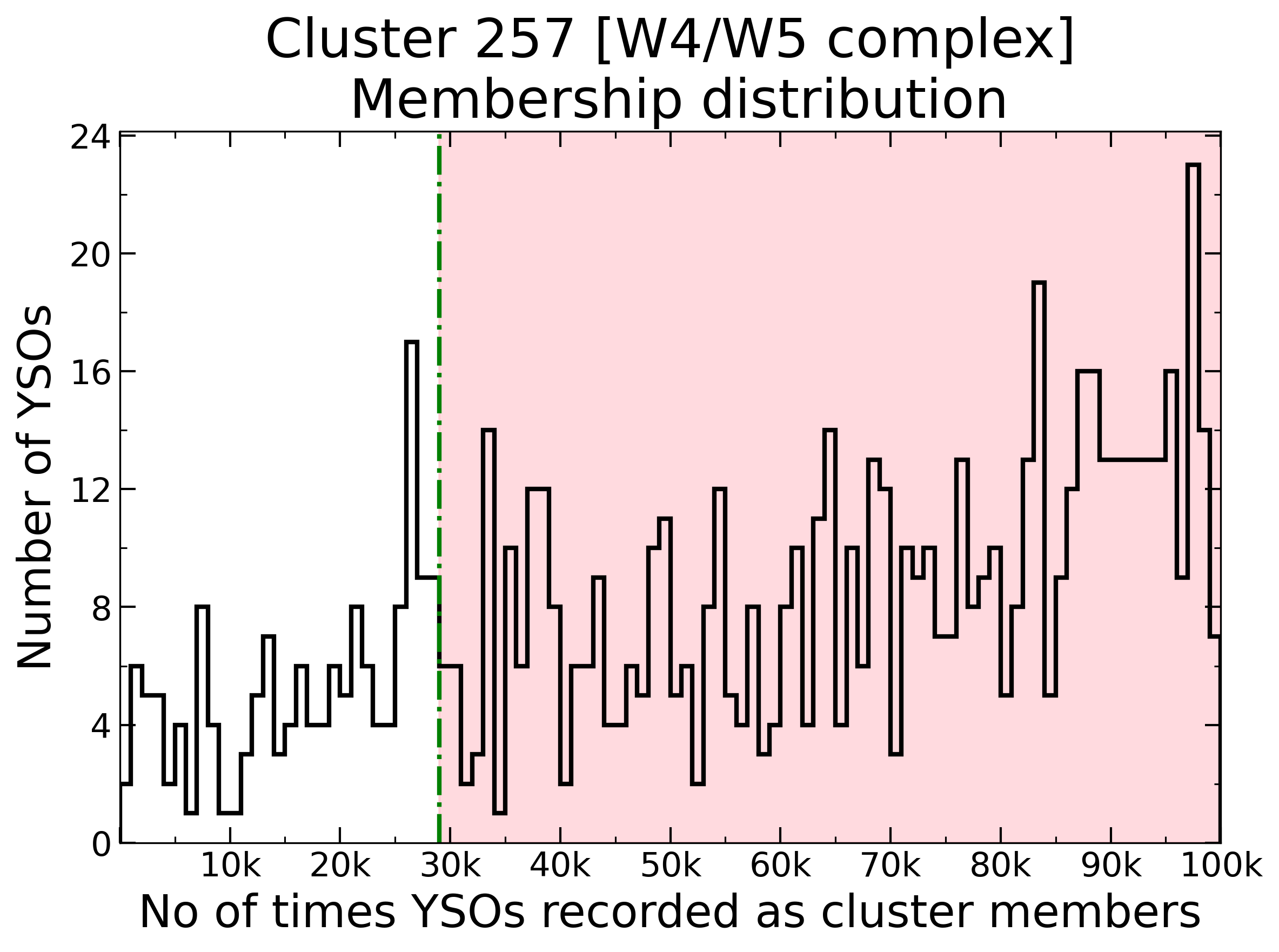} 
		\caption{Membership distribution for one of the 10 HDBSCAN-MC runs (corresponding to the horizontal axis in Figure \ref{fig: st_test_257}) generated for Cluster 257 in the W4/W5 complex. The horizontal axis indicates the number of times YSOs are identified as cluster members, while the vertical axis shows the count of such YSOs within each histogram bin. The dotted green line represents the Monte Carlo threshold ($\mathcal{MC}_{thr} = 29,000$) set for this cluster.}\label{fig: mem_dist_257}
\end{figure}

\subsection{Parallelization on High Performance Clusters \label{subsec:hpc}}

We significantly enhanced performance by parallelizing the algorithm, enabling 100,000 independent Monte Carlo iterations across available CPU cores. This parallel execution drastically reduced runtime while ensuring stable and consistent cluster memberships. Figure \ref{fig: st_test_257} compares stability over $10^{3}$ and $10^{5}$ Monte Carlo iterations for the cluster in  W4/W5 complex (Cluster 257). The Smithsonian Institution Hydra cluster was used for HDBSCAN-MC parallelization and parallelized global optimization for cluster simulations. 

\begin{figure*}[t]
 \centering
  \includegraphics[width=0.95\linewidth]{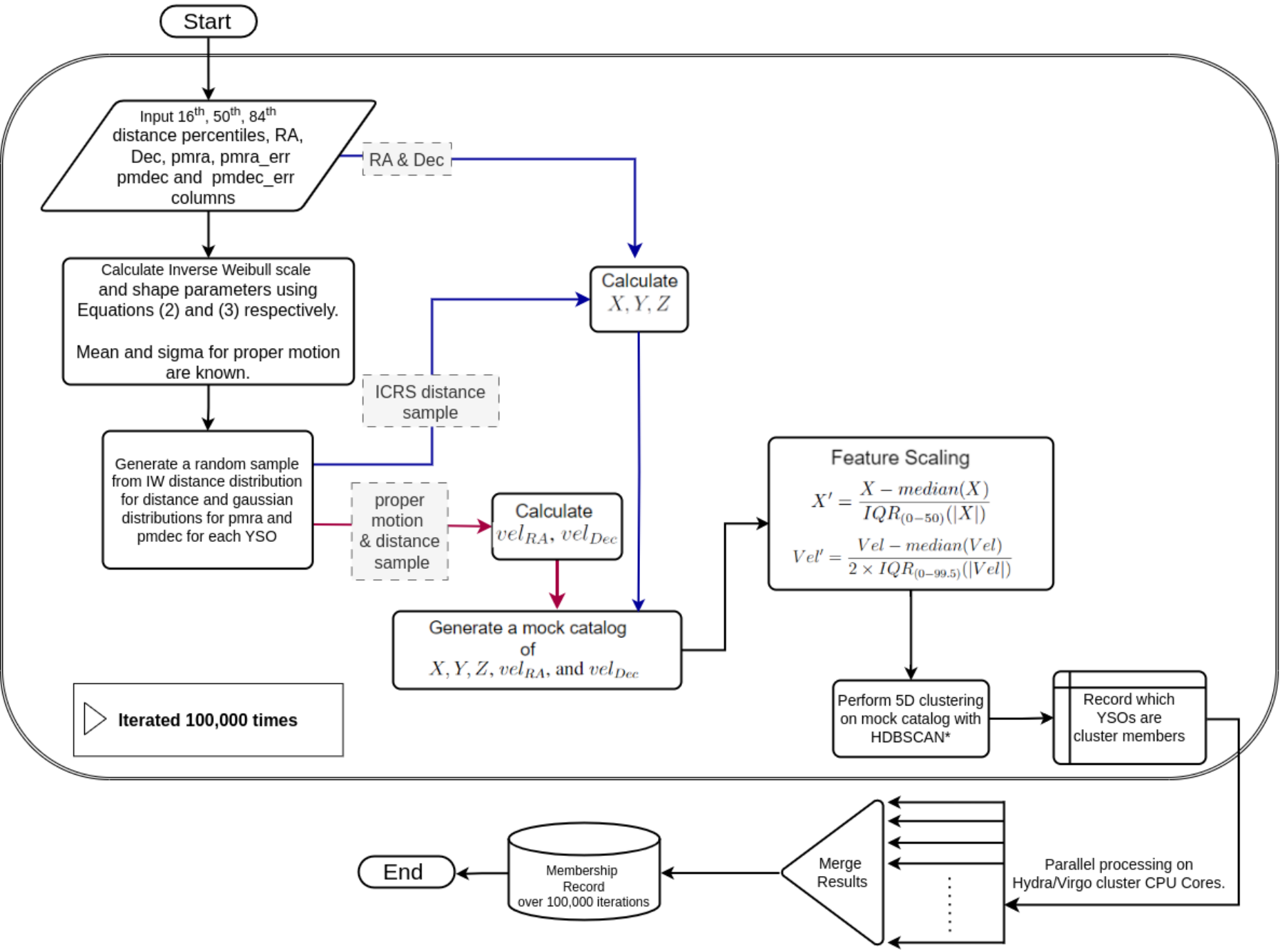} 
		\caption{High-level flowchart depicting HDBSCAN-MC execution process. To enhance readability, the arrows associated with calculation of positional (X, Y, Z) and velocity dimensions ($Vel_{RA}$, $Vel_{Dec}$) are color-coded in blue and red, while the rest of the data/information flow is depicted using black arrows. The algorithm produces the \textit{membership record array}, indicating the number of times each YSO is identified as a cluster member across $10^{5}$ Monte Carlo iterations of three-dimensional sampling. This involves generating a mock YSO cluster catalog by sampling from the ICRS distance error distribution, proper motion in Right Ascension (pmra or $\rm\mu_{\alpha} \cos\delta$), and proper motion in Declination (pmdec or $\mu_{\delta}$) for each YSO in the RA-Dec region under analysis.}
        \label{fig: HDBSCAN_MC_2}
\end{figure*}

\subsection{The Monte Carlo threshold and Need for Cluster Simulations}

The membership distribution is produced using the \textit{membership record} created by the HDBSCAN-MC algorithm. An entry in the \textit{membership record} array gives the number of iterations in which that YSO was identified as a cluster member. Once the \textit{membership record} array is created, the histogram is plotted for these numbers. The trend of this histogram is shown by the corresponding scaled kernel density estimate (KDE) (e.g., see Figure \ref{fig: mem_dist_257}). A YSO is considered a genuine cluster member if it is identified by HDBSCAN-MC as cluster members more frequently than a specified threshold, hereafter referred to as the Monte Carlo threshold ($\mathcal{MC}_{\mathrm{thr}}$). For instance, with a threshold of 29,000 (indicated by the green dash-dotted line in Figure \ref{fig: mem_dist_257}), a YSO must be identified as a member in more than 29,000 out of the 100,000 iterations to be classified as a true cluster member. The overall HDBSCAN-MC workflow is illustrated in Figure \ref{fig: HDBSCAN_MC_2}.

A simplistic approach is to set $\mathcal{MC}_{\mathrm{thr}} = 50{,}000$ for $10^{5}$ Monte Carlo iterations. However, this ad hoc approximation is unreliable, particularly for distant clusters with significant astrometric uncertainties. Determining an accurate $\mathcal{MC}_{thr}$ is challenging, as it depends on factors such as the spatial distribution of members, their proper motions, and uncertainties in Gaia distance estimates. To robustly determine $\mathcal{MC}_{\mathrm{thr}}$, we simulate clusters and field stars using the Generation Of cLuster anD FIeld STar (GOLDFIST) framework, designed to reproduce the observed characteristics of the analyzed region. By comparing the observed cluster to these simulations, where true members are known, we can determine an appropriate value for $\mathcal{MC}_{\mathrm{thr}}$ and assess the effectiveness of the HDBSCAN-MC algorithm.

\section{Generation Of cLuster anD FIeld STar (GOLDFIST) simulation} \label{sec: goldfist}

The methodology for cluster and field star simulation with GOLDFIST framework is outlined in this section. The stars in the analyzed skyplane region are classified into two distinct populations: field stars and cluster member YSOs.
\newpage
\subsection{Confirmed field YSOs}
 Certain YSOs in the analyzed region can be classified as \textit{confirmed field YSOs} if they meet one of the following criteria: (a) they are high-proper-motion stars that lie outside the $3\sigma$ range of the proper motion distribution in Right Ascension ($\mu_{\alpha} \cos\delta$) or Declination ($\mu_{\delta}$), similar to  the concept applied by \citet{dias2014proper} and suggested by \citet{jun1982discussion}; or (b) The $3\sigma$ range of the YSO's distance uncertainty, which is modeled as an inverse Weibull distribution does not overlap with the predicted cluster distance range say [$\mathcal{D}_{1}$, $\mathcal{D}_{2}$]. Hence, the \textit{confirmed field YSOs} are a function of $\mathcal{D}_{1}$, $\mathcal{D}_{2}$ and cluster's observed RA-Dec proper motion distribution.

\subsection{Parameter Space Development}
\begin{figure}[!h]
		\centering
	\includegraphics[width=0.85\linewidth]{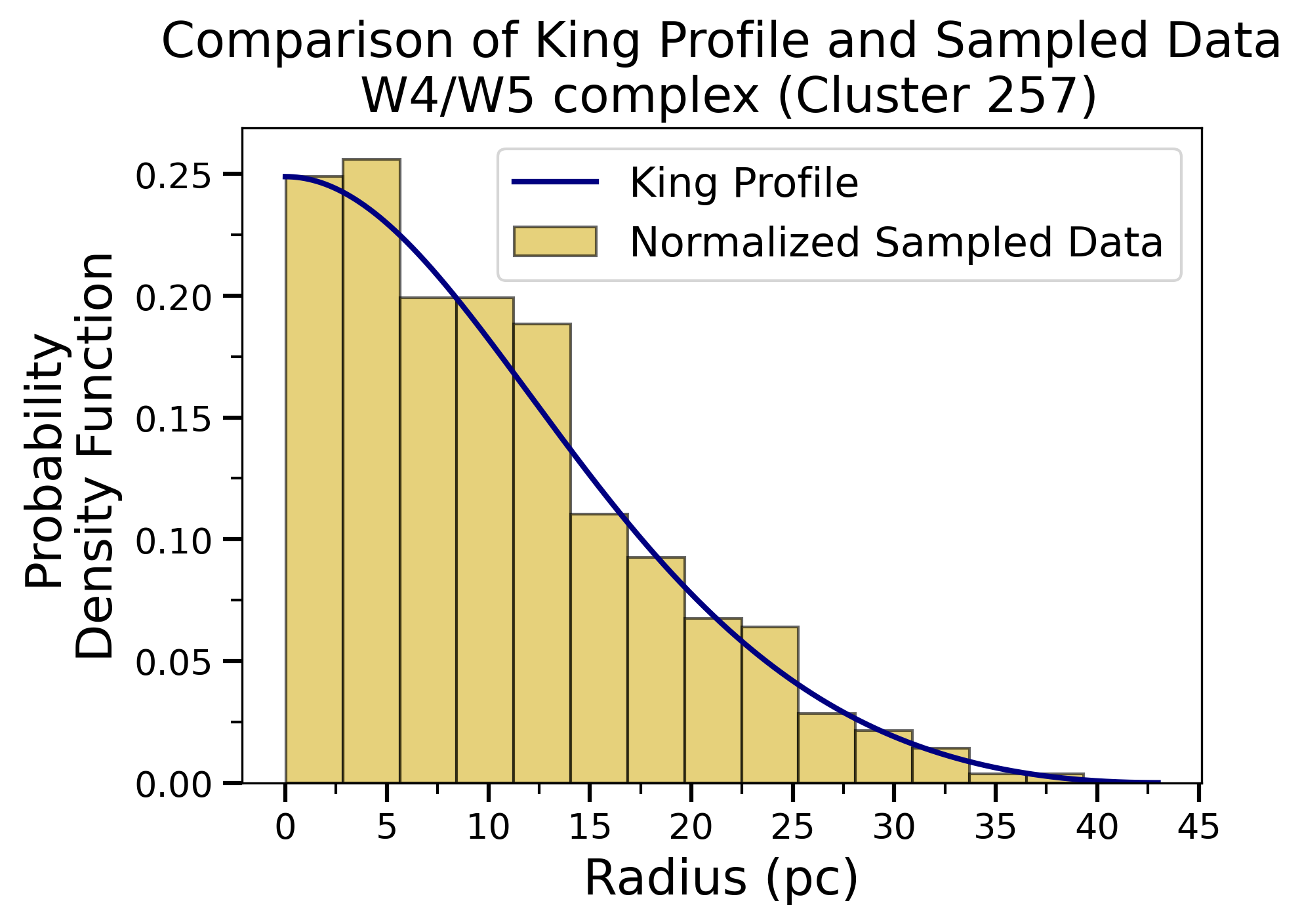} 
		\caption{Sampling distances to simulate cluster with King profile. As an example, Cluster 257 is simulated here; tidal radius, $r_t \approx 43 pc$, concentration parameter, \textbf{$\xi_{c}$} = 0.579 and \textbf{$\Phi_{mem}$} = 0.58 which corresponds to 452 cluster members out of 779 YSOs in the circular RA-Dec analysis region.}
		\label{fig: king_profile_sampling}
\end{figure}

In this work, we adopt the approximated analytic King model \citep{King1962}, as implemented in Astropy's \texttt{KingProjectedAnalytic1D} \citep{astropy:2022}, to simulate clusters. The projected surface density profile, given by Equation 14 of \cite{King1962}, captures the decline in stellar density near the cluster boundaries, where members blend into the field population. Unlike the 3D isotropic Gaussian distribution, the King profile extends to a finite tidal radius, providing better control over the cluster concentration parameter, $\xi_{c}$. The adopted King profile, although approximate, is valid for $\frac{r_c}{r_t} > 0.01$, making it suitable for embedded or open clusters analyzed in this study where $0.01 < \frac{r_c}{r_t} < 1$. Here $r_t$ refers to the tidal radius and $r_c$ represents the core radius, while the scaling factor is set to 1. The ICRS distance of the simulated cluster is specified by the \textbf{$\mathcal{D}_{center}$} parameter (in pc). The tidal radius is calculated by multiplying the tangent of the estimated cluster radius (using radius estimates from \citet{Winston2020} for the analyzed cluster) with simulated cluster distance, \textbf{$\mathcal{D}_{center}$}. The core radius is derived using the concentration parameter, \textbf{$\xi_{c}$} = $\frac{r_c}{r_t}$. As a representative example, Figure \ref{fig: king_profile_sampling} illustrates sampling from the King profile for the W4/W5 complex (Cluster 257). A fraction \textbf{$\Phi_{mem}$} of the YSOs present in the analyzed circular RA-Dec region are cluster members, while the remaining $1 - \Phi_{mem}$ fraction is classified as field YSOs. The field YSOs are uniformly distributed in the circular RA-Dec space up to a radius from the cluster center, equal to the maximum RA-Dec separation of YSOs from the cluster center as obtained from observed data. The \textbf{$F_{\mathcal{L}}$} (in pc) and \textbf{$F_{\mathcal{U}}$} (in pc) parameters define the lower and upper limits of the field YSO cone along the ICRS distance axis, respectively. These field YSOs are distributed uniformly between \textbf{$F_{\mathcal{L}}$} and \textbf{$F_{\mathcal{U}}$}, excluding the volume occupied by the cluster. Together, these parameters constitute the five spatial simulation parameters described in Table \ref{tab:parameter_table}.

The Gaia $B_{p}-R_{p}$ color and $G$-band magnitude were assigned to the simulated YSOs based on the observed cluster data, minimizing the total spherical separation in Galactic coordinates between the observed and simulated YSOs. This was implemented using a modified Jonker-Volgenant algorithm without initialization, as described in \citet{linear_sum_assignment}. The goal was to assign $B_{p}-R_{p}$ color and Gaia $G$-band magnitude in a way that carefully replicated local extinction. We therefore constructed the cost matrix based on Galactic coordinate separation to match YSOs with the least separation between the observed and simulated data.

\begin{figure*}[!ht]
		\centering
		\includegraphics[width=\linewidth]{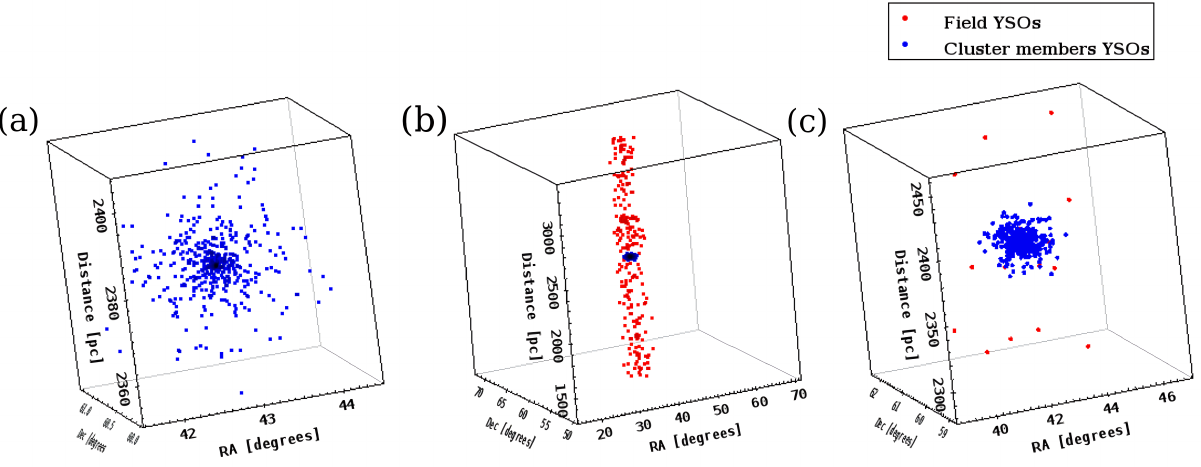} 
		\caption{Example of ideal simulation for Cluster 257 [W4/W5 complex] with King model in (RA, Dec, ICRS distance) space (a) Simulated cluster with King profile. Blue points represent cluster members. There are 452 members in the simulated YSO cluster. (b) Represents the simulated cluster embedded in a uniformly distributed low-density foreground and background field YSO cone. (c) Distribution of cluster and field YSOs plotted on nearly the same scale as in (a). About 58\% of the total YSOs (779) are cluster members (452), which implies model parameter \textbf{\textbf{$\Phi_{mem}$}} = 0.58.}
    \label{fig: king_demonstration}
\end{figure*}

This completes the spatial simulation for an ideal cluster. Figure \ref{fig: king_demonstration} presents an example of an ideal cluster simulation (Cluster 257). Appendix \ref{app: goldfist_dist} describes how the errors in the Gaia distance estimates are incorporated in the cluster simulations. 

\begin{deluxetable}{c >{\centering\arraybackslash}p{1.5cm} p{5cm}}
\tabletypesize{\scriptsize}
\tablecaption{Parameters for cluster simulation.\label{tab:parameter_table}}
\tablewidth{0pt}
\tablehead{
\colhead{\textbf{Sr. No}} & \colhead{\textbf{Parameters}} & \colhead{\textbf{Explanation}}
}
\startdata
\textbf{1} & \textbf{$\mathcal{D}_{center}$} & Controls the ICRS center distance (in pc) of the cluster modeled with the King profile. \\ \hline
\textbf{2} & \textbf{$\Phi_{mem}$} & The fraction of YSOs in the circular RA-Dec field that are cluster members. \textbf{$\Phi_{mem}$} $\in$ (0,1). \\ \hline
\textbf{3} & \textbf{$F_{\mathcal{L}}$} & The approximate ICRS distance in the cluster foreground up to which field YSOs are present. This parameter determines the lower limit of the field YSO cone along the distance axis. The foreground limit, \textbf{$F_{\mathcal{L}}$}, is typically set from the minimum of the lower uncertainty ($16^{th}$ percentile) YSO distance limit distribution to the 2.5 percentile of the upper uncertainty ($84^{th}$ percentile) distance limit distribution. \\ \hline
\textbf{4} & \textbf{$F_{\mathcal{U}}$} & The approximate ICRS distance in the cluster background up to which field YSOs are present. This parameter determines the upper limit of the field YSO cone along the distance axis. The background limit, \textbf{$F_{\mathcal{U}}$}, is typically set from the 97.5 percentile of the lower uncertainty ($16^{th}$ percentile) YSO distance limit distribution to the 99.85 percentile of the upper uncertainty ($84^{th}$ percentile) distance limit distribution. \\ \hline
\textbf{5} & \textbf{$\xi_{c}$} & Represents the concentration parameter of the cluster, derived from the King profile. \textbf{$\xi_{c}$} = $\frac{r_c}{r_t}$, where $r_{c}$ and $r_{t}$ represent the core and tidal radius in parsec respectively. \\ \hline
\textbf{6} & \textbf{$\Lambda_{pm}$} & The factor by which the simulated proper motions of the cluster members increase when extrapolated from the Taurus-Auriga complex distance to \textbf{$\mathcal{D}_{center}$}. See Equation \ref{eq: pm_scale} for details. \\ 
\enddata
\end{deluxetable}

\begin{figure}[h]
		\centering
		\includegraphics[width=\linewidth]{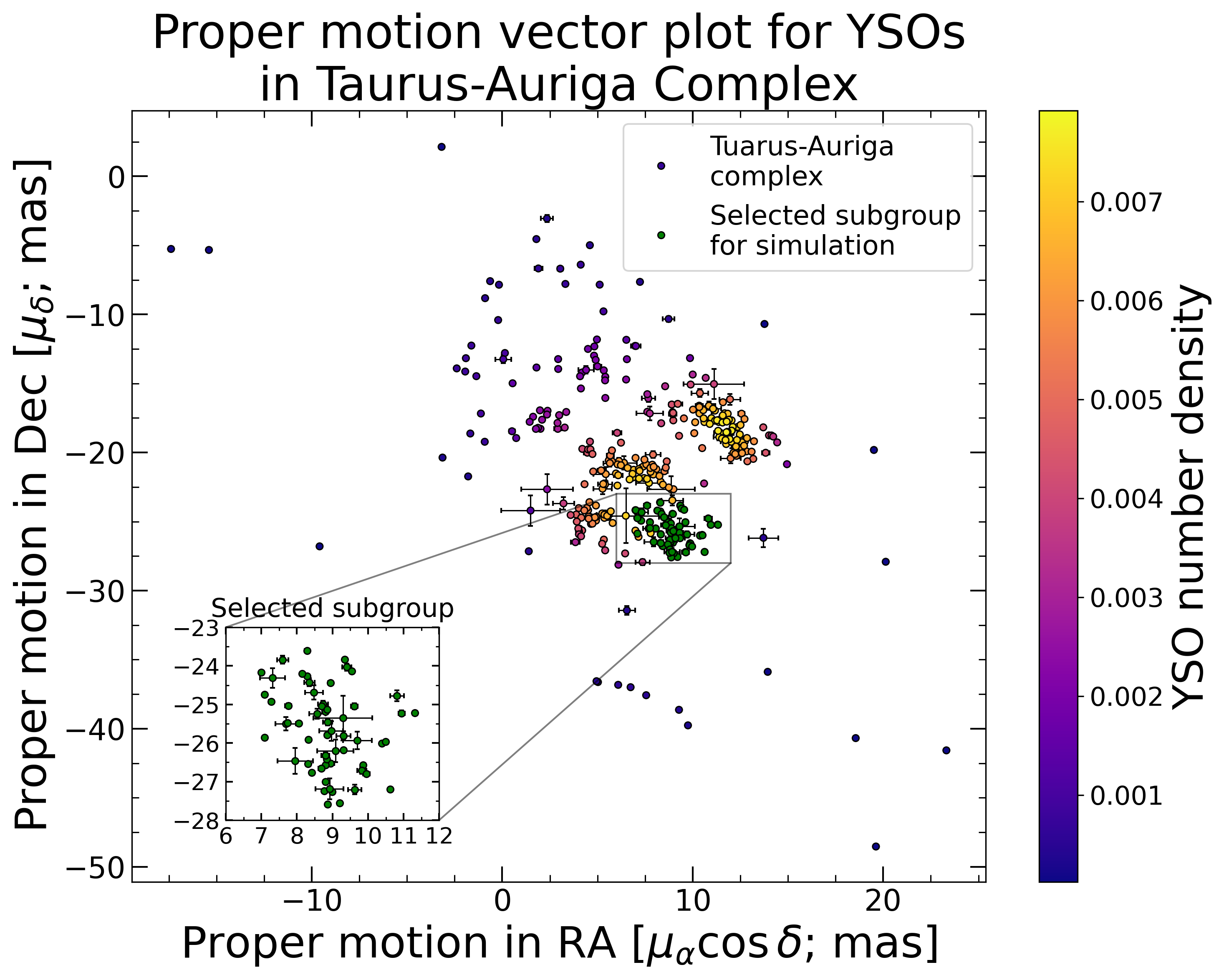} 
		\caption{Proper motion vector plot for YSOs in the Taurus-Auriga complex. Red points represent 349 YSOs with Gaia counterparts exhibiting $\texttt{ruwe} < 1.4$. The colorbar indicates the YSO number density, while the inset highlights 58 YSOs (green circles) selected to extrapolate proper motions for simulating cluster members.}
		\label{fig: TA_pm}
\end{figure}

To apply 5-dimensional clustering with HDBSCAN-MC, it was necessary to simulate the proper motions of the clusters as well. To mimic the observed variations in the proper motion vectors, we extrapolated the proper motion distribution of the sources in the Taurus-Auriga region (nearby open cluster) to the simulated cluster distance, \textbf{$\mathcal{D}_{center}$}. A total of  587 sources in the Taurus-Auriga complex with Gaia EDR3 counterparts were documented by \cite{Krolikowski2021}. We extracted the Gaia DR3 data for these sources, and selected those YSOs with a renormalized unit weight error (ruwe) < 1.4, reducing the number of reliable candidates to 349. Although distinct subgroups were observed in proper motion space, such substructures are not typically observable in distant outer Galaxy YSO clusters. Therefore we selected one of the proper motion substructures with 58 stars (see Figure \ref{fig: TA_pm}) to extrapolate the proper motions to the simulated cluster's center distance.

To generate proper motion data for all cluster members (i.e., the \textbf{$\Phi_{mem}$} fraction of total YSOs in the analyzed region) from the available 58 sources, we assumed that additional members would follow similar distributions for proper motion in Right Ascension ($\rm \mu_{\alpha}cos\delta$), Declination ($\mu_{\delta}$), and parallax, as observed in the Taurus-Auriga complex. Additional proper motion and parallax data were generated by resampling from the kernel density estimates (KDE) approximating $\rm\mu_{\alpha}cos\delta$, $\mu_{\delta}$ and parallax distributions. Scott's rule \citep{scott2015multivariate} was used to perfom the KDE approximation. We then extrapolated the proper motions of the cluster members to the \textbf{$\mathcal{D}_{center}$} of the simulated cluster using Equation set \ref{eq: pm_scale}.

\begin{align}
    \begin{split}
    \rm\mu_{\alpha}cos\delta^{\dag} &= \text{\textbf{$\Lambda_{pm}$}} \times \frac{1000 \times \text{TA}_{\rm \mu_{\alpha}cos\delta}}{\text{TA}_{\rm parallax} \times \text{\textbf{$\mathcal{D}_{center}$}}} \\
    {\mu_{\delta}}^{\dag} &= \text{\textbf{$\Lambda_{pm}$}} \times \frac{1000 \times \text{TA}_{\mu_{\delta}}}{\text{TA}_{\rm parallax} \times \text{\textbf{$\mathcal{D}_{center}$}}}
    \end{split}
    \label{eq: pm_scale}
\end{align}

The $\rm\mu_{\alpha}cos\delta^{\dag}$ and ${\mu_{\delta}}^{\dag}$ in Equation set \ref{eq: pm_scale} refer to the extrapolated proper motions for simulated cluster members in RA and Dec, respectively. $\text{TA}_{\rm\mu_{\alpha}cos\delta}$, $\text{TA}_{\mu_{\delta}}$, and $\text{TA}_{\rm parallax}$ represent the proper motion in RA, Dec, and parallax for the Taurus-Auriga (TA) complex members which includes additional YSOs generated through KDE sampling. The parameter \textbf{$\Lambda_{pm}$} controls the scaling during the extrapolation of proper motions from the TA complex to the simulated YSO cluster. The factor of 1000 accounts for the milliarcsecond units of $\text{TA}_{\rm parallax}$. Two-dimensional KDE fitting estimated the peak of the proper motion distribution in the observed data. Subsequently, we shifted the median of the extrapolated proper motions for the cluster members to align with this peak, ensuring a more accurate representation of the observation. Hence, the extrapolated ($\rm\mu_{\alpha}cos\delta^{\dag}$, ${\mu_{\delta}}^{\dag}$), followed by shifting of the median, served as the mean proper motions for the simulated cluster members. The procedure for assigning proper motion uncertainties to the simulated YSOs is detailed in Appendix \ref{app: goldfist_prop}. The final simulated mean proper motions are henceforth denoted as ($\rm\mu_{\alpha}cos\delta_{sim}$, ${\mu_{\delta}}_{\rm sim}$).

The simulation description for the specific parameter set, summarized in Table \ref{tab:parameter_table}, is now complete. Table \ref{tab:parameter_table} summarizes the six parameters used for cluster simulations along with their descriptions.

\begin{figure}[htpb]
		\centering
		\includegraphics[width=\linewidth]{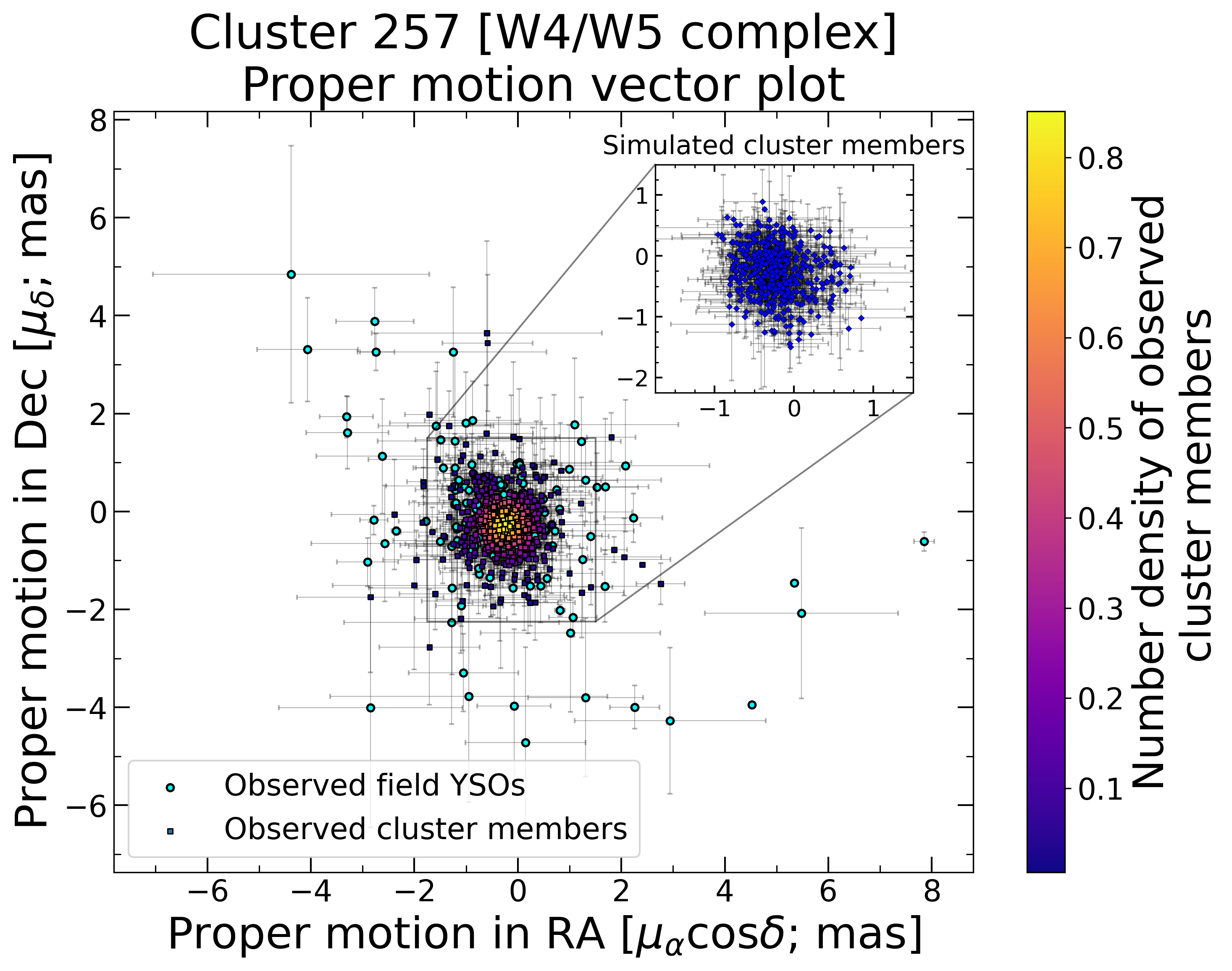} 
		\caption{Proper motion vector plot of the Cluster 257 (W4/W5 complex), comparing observed and simulated cluster members. The color scale represents the number density of observed cluster members (depicted as squares) identified after applying the simulation-predicted Monte Carlo threshold ($\mathcal{MC}_{thr}$). Observed field YSOs are marked by cyan circles. The inset highlights the simulated cluster members for clarity. Grey error bars, displayed with reduced opacity, indicate proper motion uncertainties. This figure illustrates the comparison between the proper motions of simulated and observed cluster members in the W4/W5 region.}
		\label{fig: c257_full_pm}
\end{figure}

Figure \ref{fig: c257_full_pm} shows the proper motion vector plot for the W4/W5 complex (Cluster 257), including both observed and simulated YSOs. The application of $\mathcal{MC}_{thr}$, determined through global optimization, allowed us to identify the cluster members displayed in the plot. This optimization, as detailed in the next section, aimed to vary the values of the six simulation parameters to find a simulation that closely matched the observed spatial and proper motion distributions of YSOs.

The simulated field YSOs, will exceed the number of confirmed field YSOs due to the spatial optimization constraint (see section \ref{sec: optimization}). At the end of each optimization iteration, few field YSOs in simulation were replaced with the observation-based closest confirmed field YSOs corresponding to the predicted cluster distance range [\textbf{$\mathcal{D}_{center}$} - $r_t$, \textbf{$\mathcal{D}_{center}$} + $r_t$]. This replacement was again carried out using the modified Jonker-Volgenant assignment, with a cost matrix based on the differences between the observed median and final simulated field YSO distances. This update, represented by the green square in the high-level block diagram in Figure \ref{fig: block_optimization}, served as a correction step, refining the optimization for a more accurate and reliable solution.

\subsection{Simulating YSO Clusters with Genetic Optimization}\label{sec: optimization}

The objective is to simulate a cluster that closely matches the spatial 3D distribution of YSOs and proper motion distributions in RA and Dec by optimizing the parameters given in Table \ref{tab:parameter_table}. Such a cluster is anticipated to closely match the observation.

Differential Evolution (DE) was used for Stage-1 and Stage-2 optimization, as described in subsections \ref{subsec: spatial_optimization} and \ref{subsec: pm_optimization}. Introduced by \citet{storn1995differential}, DE is a global optimization algorithm inspired by natural selection. As a metaheuristic method, DE requires minimal assumptions and primarily focuses on parameter bounds. Its ability to efficiently navigate large solution spaces makes it well-suited for the multidimensional global optimization required in this study's automated cluster simulations.

Before performing cluster simulation and optimization, we removed the confirmed field YSOs corresponding to the cluster distance bounds used in the optimization. This step simplified the simulation process and improved optimization efficiency.

\subsubsection{Stage-1: Optimizing spatial YSO distribution to match observed distribution.}\label{subsec: spatial_optimization}

To simulate a cluster closely resembling the observed one, the initial step involves aligning the spatial probability distribution of observed YSOs with that of the simulated cluster. Here, the Galactocentric coordinates—X, Y, and Z—represent the spatial positions of the YSOs within the analyzed region. Likewise, the coordinates x, y, and z correspond to the Galactocentric positions of the simulated YSOs. Apart from the spatial coordinates, the objective function also incorporates their probability distribution and the Kullback–Leibler (KL) divergence. The probability distribution, for example, of the X coordinate, $\mathds{P}(\mathrm{X})$, is computed as $\text{hist}(\mathrm{X})/L$, where $\text{hist}(\mathrm{X})$ denotes the histogram frequencies of the X positions and $L$ is the total number of YSOs in the analyzed region. Further details on the evaluation of coordinate probability distribution are provided in Appendix \ref{app: goldfist_spatial_opt}.

The KL divergence, or relative entropy, measures the difference between a probability distribution and a reference distribution, quantifying the information loss when one distribution approximates the other. A value of zero indicates identical distributions, making KL divergence an effective candidate for constructing an objective function to match simulated cluster distributions to observed ones. Accordingly, the objective function defined in Equation \ref{eq:f} is designed to simultaneously align the probability distributions of X, Y, and Z with their corresponding simulated coordinate sets, x, y, and z. 

The term `KL' represents KL divergence adapted for convex optimization (see \citealt{Boyd2004}) in Equation \ref{eq:f}. 

\vspace{-1mm}
\begin{equation}
    F(X,Y,Z,x,y,z) = \sum_{i=1}^{3} \text{KL}\{\mathds{P}(X_{i}), \mathds{P}(x_{i})\}\label{eq:f}
\end{equation}

\begin{align}
\begin{split}
    \Phi_{mem} &< 1 - \frac{\mathcal{N}_{\text{confirmed field YSOs}}}{\mathcal{N}_{\text{total}}} \\
    F_{\mathcal{L}} &< \mathcal{D}_{\text{center}} < F_{\mathcal{U}}
\end{split}
\quad  \Bigg\} \text{ Constraints} 
\label{eq:f_constraints}
\end{align}

\vspace{2pt}

The implementation utilizes the `scipy.special.kl\_div' module \citep{2020SciPy-NMeth}. Here, $X_{i}$ (for \( i = 1 \) to \( 3 \)) denotes observed Galactocentric coordinates \( X, Y, \) and \( Z \), and similarly, \( x_{i} \) (for \( i = 1 \) to \( 3 \)) represents simulated Galactocentric coordinates \( x, y, \) and \( z \). The generation of spatial coordinates x, y, and z of the simulated cluster are determined by the first 5 parameters as mentioned in Table \ref{tab:parameter_table}. Namely \textbf{$\mathcal{D}_{center}$}, \textbf{$\Phi_{mem}$}, \textbf{$F_{\mathcal{L}}$}, \textbf{$F_{\mathcal{U}}$}, and concentration parameter \textbf{$\xi_{c}$}. The constraints used for spatial optimization are given by inequalities \ref{eq:f_constraints}. In an optimization iteration,  $\mathcal{N}_{\text{confirmed field YSOs}}$ depicts the number of confirmed field YSOs corresponding to the predicted cluster distance range [\textbf{$\mathcal{D}_{center}$} - $r_t$, \textbf{$\mathcal{D}_{center}$} + $r_t$], while $\mathcal{N}_{\text{total}}$ denotes the total number of YSOs in the analyzed circular RA-Dec region. $\mathcal{N}_{\text{total}}$ remains constant throughout all iterations. The final paragraph of Appendix \ref{app: goldfist_spatial_opt} outlines the determination of parameter bounds for spatial optimization.

\subsubsection{Stage-2: Optimizing proper motion distribution to match observed YSO proper motions.}\label{subsec: pm_optimization}

Once the spatial optimization is completed the next step is to simulate the YSO proper motions. This requires the five optimized parameters from Stage-1 to get optimum values for proper motion scaling parameter \textbf{$\Lambda_{pm}$}. The objective function (G) for proper motion optimization is given by Equation \ref{eq:G}.

In Equation \ref{eq:G}, $j=\{1,2\}$ refers to the observed proper motions ($\rm\mu_{\alpha}cos\delta$, $\mu_{\delta}$) and the simulated proper motions ($\rm\mu_{\alpha}cos\delta_{sim}$, ${\mu_{\delta}}_{\rm sim}$), respectively.

\begin{equation}
    G = \sum_{j=1}^{2} \text{KL}\{\mathds{P}(\rm\mu_{\alpha}cos\delta_{j}), \mathds{P}({\mu_{\delta}}_{\rm j})\}\label{eq:G}
\end{equation}

\begin{figure}[!h]
		\centering
\includegraphics[width=1.1\linewidth]{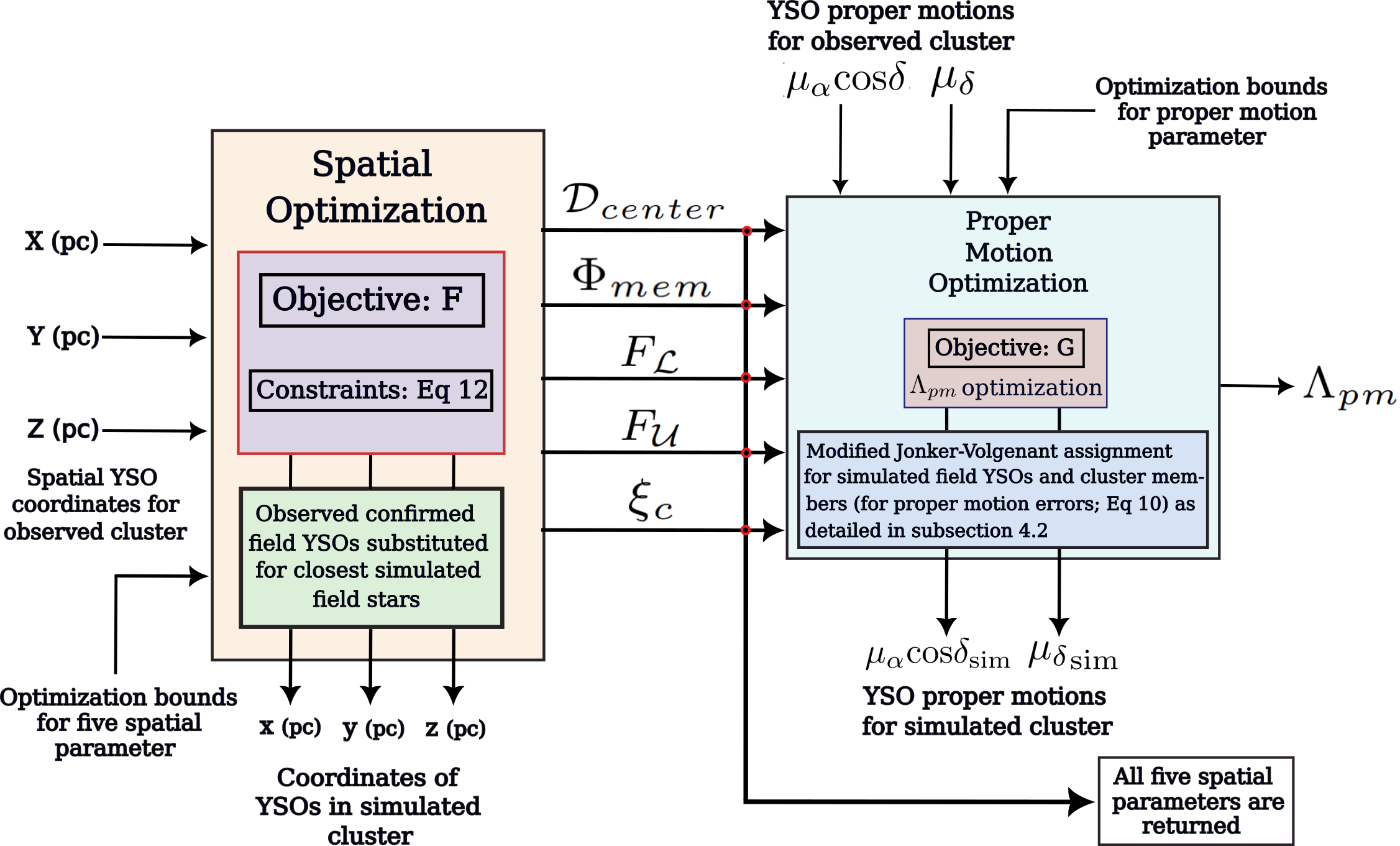} 
    \caption{High-level block diagram illustrating cluster simulation using multidimensional global optimization. The diagram illustrates a single iteration of the optimization process. For both spatial and proper motion optimizations, \textit{Popsize} in differential evolution is fixed at 150, yielding a total population ($N_p$) of 750 for five-parameter and 150 for single-parameter optimization. The mutation factor (F) varies between 0.5 and 1.5, with the crossover probability ($P_{cr}$) set at 0.5. Both optimizations employ the 'best1bin' strategy with a tolerance (\textit{tol}) of 0.5. Computations are parallelized across 80 CPU cores for enhanced computational efficiency during both stages of optimization.}\label{fig: block_optimization}
\end{figure}

The calculation of the probability distribution $\mathds{P}(\mu)$ for Equation \ref{eq:G} follows the same approach as described for spatial optimization in subsection \ref{subsec: spatial_optimization}. Similarly, as in Stage-1 optimization, the objective function $G$ (Equation \ref{eq:G}) is minimized using Differential Evolution. The proper motion parameter, \textbf{$\Lambda_{pm}$}, is constrained within 0.1 to 9.9. The entire optimization process is repeated 60 times, and the optimized parameter sets are sorted in ascending order of the objective function values. It is anticipated that parameter sets yielding a close match to the 3D spatial distribution and the proper motion distributions in RA and Dec are promising candidates for simulating cluster close to the observation. The block diagram in Figure \ref{fig: block_optimization} provides a quick overview of the optimization process for a single iteration.

Once the cluster members are identified, those identified as confirmed field YSOs, within the distance range $[\text{Min}(\mathcal{D}_{\text{center; opt}}) - r_t, \text{Max}(\mathcal{D}_{\text{center; opt}}) + r_t]$, are excluded from the analyzed cluster membership. Here, $\mathcal{D}_{\text{center; opt}}$ denotes the final simulation-predicted optimized cluster distances (see Appendix \ref{app: DE_tables}, Table \ref{tab: params}).

\begin{figure*}[t]
\centering
    \includegraphics[width=0.71\linewidth]{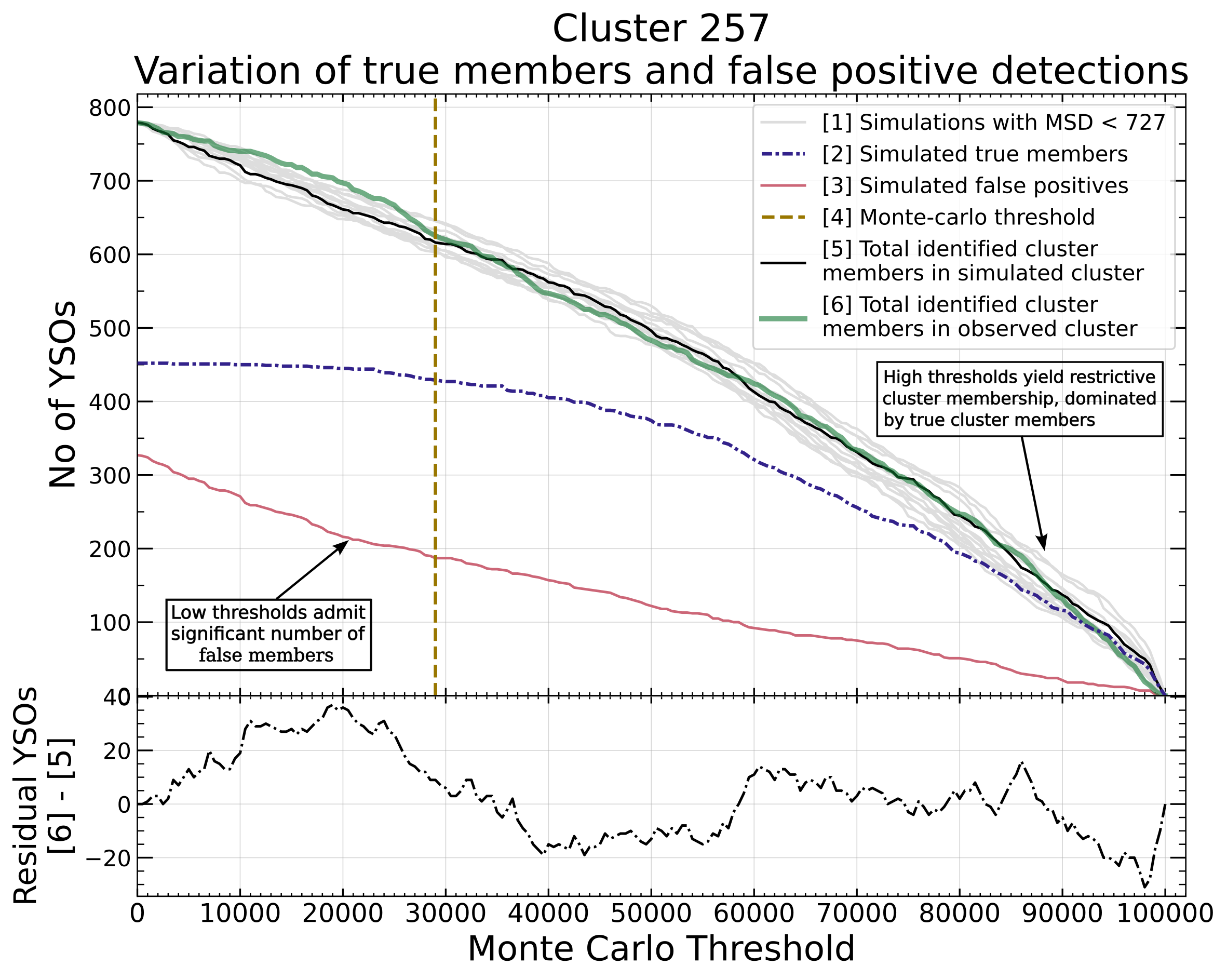} 
    \caption{Interpreting complementary cumulative plot (plotting number of retained YSOs on increasing $\mathcal{MC}_{thr}$), henceforth referred to as the Monte-Carlo spectrum, for the field and member YSOs of the W4/W5 complex (Cluster 257). The kinematic distance estimate for this cluster is 2000 pc. The green curve illustrates the decrease in observed total cluster members with increasing threshold. The black curve represents the identified total cluster members for the simulated cluster. The blue dash-dotted line and red curve depict the decline in simulated true members and false positives, respectively, with increasing threshold. The difference between identified total cluster members for the observed and simulated clusters is shown by the black dash-dotted line in the residual plot. The faded grey lines represent the total identified cluster members for simulated clusters generated using an alternative set of parameters, with mean squared deviation (MSD) within three times the MSD of the best-fit black curve from the green plot. The yellow vertical dash-dotted line represents the final determined Monte-Carlo threshold ($\mathcal{MC}_{thr} = 29,000$ for this cluster), where the simulated true members comprise 95\% of the initial count.}
    \label{fig: mc_spectra}
\end{figure*}


\section{The Monte-Carlo Spectrum and Monte-Carlo Threshold Determination.}

A membership distribution similar to Figure \ref{fig: mem_dist_257} can also be generated for the simulated cluster, where the exact classification of YSOs as cluster members or field YSOs is known. For a particular Monte Carlo threshold, it was observed that 100\% retrieval of actual cluster members was not possible due to large distance uncertainties, leading to identification of some false positive YSOs as cluster members. The Monte Carlo threshold is varied from 0 to 100,000 to analyze how identified true cluster members and false positives vary with the threshold, generating a complementary cumulative plot. This complementary cumulative plot, or survival plot, is henceforth referred to as the \textit{Monte Carlo spectrum}. It is shown by the green and black curves in Figure \ref{fig: mc_spectra}. The best-fit black line in Figure \ref{fig: mc_spectra} represents the identified total cluster members in the simulation, revealing two distinct components in the Monte Carlo spectrum of the observed cluster (depicted by the green line). These components represent variations in true and false positives, which, when combined, contribute to the identified total cluster members. The Monte Carlo spectrum, encapsulates information on surrounding field stars, cluster kinematics, and morphology. Consequently, the final Monte Carlo threshold is determined such that the true members constitute 95\% of their original count. Sacrificing a small percentage of true members helps to significantly reduce false positives in most cases.  

Monte Carlo spectra are calculated for the top 30 optimized parameter sets, and the Mean Squared Deviation (MSD) is recorded for each case with respect to the observed Monte Carlo spectra. The parameter set with the minimum MSD ($MSD_{\text{min}}$) is declared as the best-fit solution and is used to determine the Monte Carlo threshold (best-fit solution represented by the black curve and final Monte Carlo threshold indicated by the yellow vertical line in Figure \ref{fig: mc_spectra}). The parameter sets having $MSD < 3 \times MSD_{min}$ are used to determine the weighted average parameters and the weighted unbiased standard deviation \citep{gnu_scientific_library} with weights = $\frac{1}{\text{MSD}}$. Tables \ref{tab: params} in Appendix \ref{app: DE_tables} lists parameter sets having $MSD < 3 \times MSD_{min}$, average parameter values, and their unbiased standard deviation.

The membership probability, $P_{i}$, is calculated using Equation \ref{eq:mem_prob} for a single HDBSCAN-MC run. In Equation \ref{eq:mem_prob}, $n_{i}$ represents the number of times the $i^{\text{th}}$ YSO is identified as a cluster member across $10^{5}$ Monte Carlo iterations, and $Max(n_{i})$ is the maximum $n_{i}$ value among all YSOs in the analyzed RA-Dec region.

\begin{equation}
    P_{i} = \frac{n_{i}}{Max(n_{i})}
    \label{eq:mem_prob}
\end{equation}

Table \ref{tab: cluster_tables} in Appendix \ref{app: cluster_tables} provides the average probability across 10 HDBSCAN-MC runs (see Figure \ref{fig: st_test_257}) and the associated membership probability uncertainty (standard deviation in 10 runs) for each source as the \texttt{mem\_prob} and \texttt{mem\_prob\_err} columns, respectively. The table includes data for the clusters analyzed in the following section. The Zenodo record \citet{patel2025} hosts the Python-based code developed in this work for implementing HDBSCAN-MC and the GOLDFIST simulation framework.

\section{Proof of Concept: Evaluating Methodology Across Clusters of Varied Distances and Densities}\label{sec: poc}

To assess the performance of the technique, we tested it on three clusters at different distances and of diverse densities. The simulation results for W4/W5 complex (Cluster 257), Cluster 123, and Cluster 163 are summarized in Figures \ref{fig: mc_spectra}, \ref{fig: cummulative_Cluster_123_163}(a), \ref{fig: cummulative_Cluster_123_163}(b) respectively. In each case, the best-fit black curve serves as an effective solution and representation of the actual cluster. The chosen Monte Carlo thresholds, marked by the yellow dash-dotted line reduce false positives while retaining 95\% of true members. Table \ref{tab: cluster_bounds} summarizes the parameter bounds for the analyzed clusters. The parameter sets with $MSD < 3 \times MSD_{\text{min}}$ from 30 runs that achieved the best match during spatial optimization, along with the corresponding weighted parameters, are summarized in Appendix \ref{app: DE_tables}, Table \ref{tab: params}, for the analyzed clusters. 

\subsection{Dense nearby cluster within 1000 pc}

 The kinematic distance estimate for the Cluster 123 is not known, hence distance bounds for optimization are determined as explained in Appendix \ref{app: goldfist_spatial_opt} with m = 2, n = 2 (Table \ref{tab: cluster_bounds}). Figure \ref{fig: cluster_results}(a) shows the spatial distribution of cluster members in the RA-Dec plane, overlaid on the WISE 12$\mu$m mosaic. Notably, no Gaia-matched SFOG YSOs are detected as cluster members in this field. The SFOG-only YSOs, as identified by \citet{Winston2020}, are offset from the previously determined cluster center but exhibit a strong association with the molecular cloud structure. The cluster is relatively young as most cluster members are optical Gaia-only YSOs. We also consider SFOG-only YSOs within the $30^{th}$ percentile density contour of the cluster members identified by the HDBSCAN-MC and GOLDFIST simulations as part of the cluster. However, no such YSO was found within the contour (which would appear as a bright green square, if present) in Figure \ref{fig: cluster_results}(1a). Figure \ref{fig: cluster_results}(1b) presents the ICRS distance distribution for the identified Gaia-only and Gaia-matched SFOG YSOs. The median $50^{th}$ percentile distance, 786 pc $\pm$ 75 pc is in close agreement with the simulation-predicted distance of 740 pc $\pm$ 6 pc, as summarized in Table \ref{tab: cluster_statistics}.

\subsection{Cluster in W4/W5 complex at 2000 pc}

\begin{figure}[htpb!]
\centering
\includegraphics[width=\linewidth]{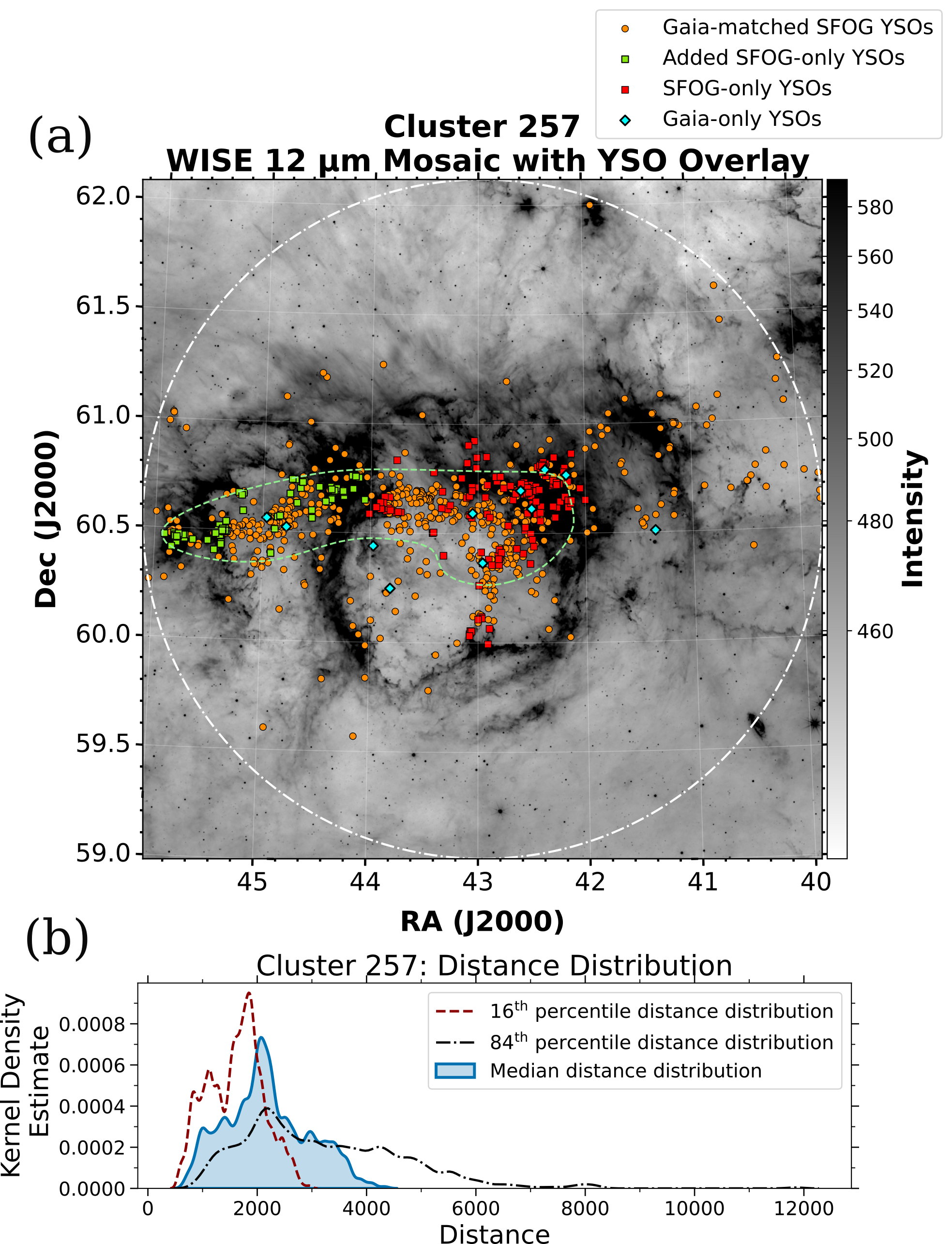} 
    \caption{Spatial distribution of cluster members in the W4/W5 complex (Cluster 257). \textbf{(a)} The subplot shows YSO cluster members overlaid on a WISE $12\mu$m inverse grayscale mosaic in the RA-Dec plane. The white dash-dotted circle indicates the region analyzed for cluster membership. Orange squares mark Gaia-matched SFOG YSOs, cyan diamonds indicate Gaia-only YSOs, and red squares denote SFOG-only YSOs from \citet{Winston2020}. SFOG-only YSOs within the $30^{th}$ percentile YSO density contour (green dashed curve) are included as cluster members (bright green squares). The contour is based on HDBSCAN-MC and GOLDFIST-identified cluster members—more evolved Gaia-only or Gaia-matched SFOG YSOs. \textbf{(b)} The ICRS distance distribution of cluster members is illustrated. The blue-filled plot shows the median distance, with the $16^{th}$ and $84^{th}$ percentiles marked by red dashed and black dash-dotted lines. This figure presents the spatial distribution of YSO cluster members in both RA-Dec and ICRS distance space for the W4/W5 region.}
\label{fig: c257_full}
\end{figure}

The kinematic distance estimate for the cluster in W4/W5 complex (Cluster 257) is 2000 pc. The distance bounds for optimization are defined to be within 20\% of the kinematic estimate. Figure \ref{fig: c257_full}(a) illustrates the spatial distribution of cluster members in the RA-Dec plane, with YSOs overlaid on the WISE 12 $\mu$m mosaic. Notably, the SFOG-only (IR) YSOs, which are less evolved compared to Gaia-matched SFOG or Gaia-only (optical) YSOs, show a stronger association with the molecular cloud structure. This is expected, as they have had less time to migrate from their formation sites compared to the more evolved cluster members that are detectable in optical bands. Given this, the SFOG-only YSOs within the $30^{th}$ percentile density contour of the cluster members identified by the HDBSCAN-MC and GOLDFIST simulations are also considered part of the cluster. These are represented by bright green squares within the contour (outlined by the green dashed closed curve) in Figure \ref{fig: c257_full}(a). The ICRS distance distribution of the identified Gaia-only and Gaia-matched SFOG YSOs is presented in Figure \ref{fig: c257_full}(b). The median of the $50^{th}$ percentile distance is 2108 pc $\pm$ 798 pc, which closely agrees with the simulation-predicted distance of 2158 pc $\pm$ 180 pc, as summarized in Table \ref{tab: cluster_statistics}. 

\begin{splitdeluxetable*}{lccccBccccBccc}
\tabletypesize{\scriptsize}
\tablewidth{0pt} 
\tablecaption{Sky positions, distance statistics, and membership records for the analyzed outer Galaxy clusters. \label{tab: cluster_statistics}}
\tablehead{
\colhead{Cluster ID} & \colhead{RA (in deg)}& \colhead{Dec (in deg)} & \colhead{Radius (in deg)} &
\colhead{2D DBSCAN} &
\colhead{5D HDBSCAN-MC} & \colhead{False Positives} & \colhead{New members} & \colhead{Outliers} & \colhead{Median Distance} & \colhead{Kinematic Estimate} & \colhead{Simulation Predicted Distance}
} 
\colnumbers
\startdata  
{Cluster 123} & 312.957906 & 44.541232 &  0.956887 &  175 & \makecell{\\ 196 out of 433  \\ Gaia-matched SFOG: 0 out of 33 \\ Gaia only: 196 out of 400}
 & $28_{-1}^{+55}$  & \makecell{\\ 196 \\ Gaia-matched SFOG: 0 \\ Gaia-only: 196 \\ SFOG-only: 0}  &  1  & 786 pc $\pm$ 75 pc &  N/A  & 740 pc $\pm$ 6 pc \\ \cline{6-6} \cline{8-8}
{Cluster 257}& 42.959137 & 60.566928 & 1.553079  &  632 & \makecell{\\ 616 out of 779 \\ Gaia-matched SFOG: 605 out of 761 \\ Gaia only: 11 out of 18}  & $187_{-38}^{+56}$  & \makecell{\\ 416 \\ Gaia-matched SFOG: 324 \\ Gaia-only: 11 \\ SFOG-only: 81} & 50  & 2108 pc $\pm$ 798 pc    & 2000 pc $\pm$ 400 pc & 2158 pc $\pm$ 180 pc\\ \cline{6-6} \cline{8-8}
{Cluster 163}& 319.954761  & 51.908440 & 1.217858   & 108  & \makecell{\\ 135 out of 201 \\ Gaia-matched SFOG: 134 out of 198 \\ Gaia-only: 1 out of 3} & $108_{-18}^{+3}$ & \makecell{\\ 142 \\ Gaia-matched SFOG: 131 \\ Gaia-only: 1 \\ SFOG-only: 10}   &  2  & 3656 pc $\pm$ 1230 pc & 9900 pc $\pm$ 1980 pc & 3531 pc $\pm$ 444 pc \\
\enddata
\tablecomments{Right Ascension (RA) and Declination (Dec) refer to the central coordinates of the cluster. The listed \textit{Radius} indicates the angular distance used as the radial cut in the sky plane, centered on the given RA and Dec, for analyzing the cluster. The values in the second table partition (columns 6,7,8 and 9) are derived from the best-fit simulation to the observed Monte Carlo spectra (see Figures \ref{fig: mc_spectra}, \ref{fig: cummulative_Cluster_123_163}(a), and \ref{fig: cummulative_Cluster_123_163}(b)). The subscripts and superscripts for the estimated false positive cluster members indicate the minimum and maximum values, respectively, based on solutions with a mean squared deviation (MSD) within three times the MSD of the best-fit Monte Carlo spectrum (black curve) from the observed Monte Carlo spectrum (solid green line) in Figures \ref{fig: mc_spectra}, \ref{fig: cummulative_Cluster_123_163}(a), and \ref{fig: cummulative_Cluster_123_163}(b). Clusters 123 and 257 may contain members from other clusters identified by \citet{Winston2020}, but since they are associated with Clusters 257 and 163, they are considered new members of these clusters. Central RA and Dec cluster coordinates are obtained from \citet{Winston2020} for Clusters 257 and 163.}
\end{splitdeluxetable*}

\begin{deluxetable*}{ccccc}
\tabletypesize{\scriptsize}
\tablecaption{Optimization bounds for Clusters 123, 257 and 163.\label{tab: cluster_bounds}}
\tablehead{
\colhead{\textbf{Sr. No}} & \colhead{\textbf{Parameters}} & \colhead{\textbf{Cluster 123}} &  \colhead{\textbf{Cluster 257}} &  \colhead{\textbf{Cluster 163}}
}
\startdata
\textbf{1} & \textbf{$\mathcal{D}_{center}$} & [673 pc , 953 pc] & [1600 pc , 2400 pc] & [2970 pc , 5940 pc] \\
\textbf{2} & \textbf{$\Phi_{mem}$} & [0.5, 0.99] & [0.5, 0.99] & [0.1, 0.99]\\
\textbf{3} & \textbf{$F_{\mathcal{L}}$} & [479.78 pc, 647.52 pc] & [269.91 pc, 1110.50 pc] & [589.70 pc, 2202.63 pc] \\
\textbf{4} & \textbf{$F_{\mathcal{U}}$} & [978.79 pc, 1800.00 pc] & [3047.75 pc, 11818.90 pc] & [5711.32 pc, 13652.26 pc] \\
\textbf{5} & \textbf{$\xi_{c}$} & [0.01, 0.99] & [0.01, 0.99] &[0.01, 0.99] \\
\textbf{6} & \textbf{$\Lambda_{pm}$} & [0.1, 9.9] & [0.1, 9.9] &[0.1, 9.9] \\
\enddata
\end{deluxetable*}


\subsection{Sparse cluster beyond 3500 pc}

\begin{figure*}[h!]
	\centering
        \includegraphics[width=\linewidth]{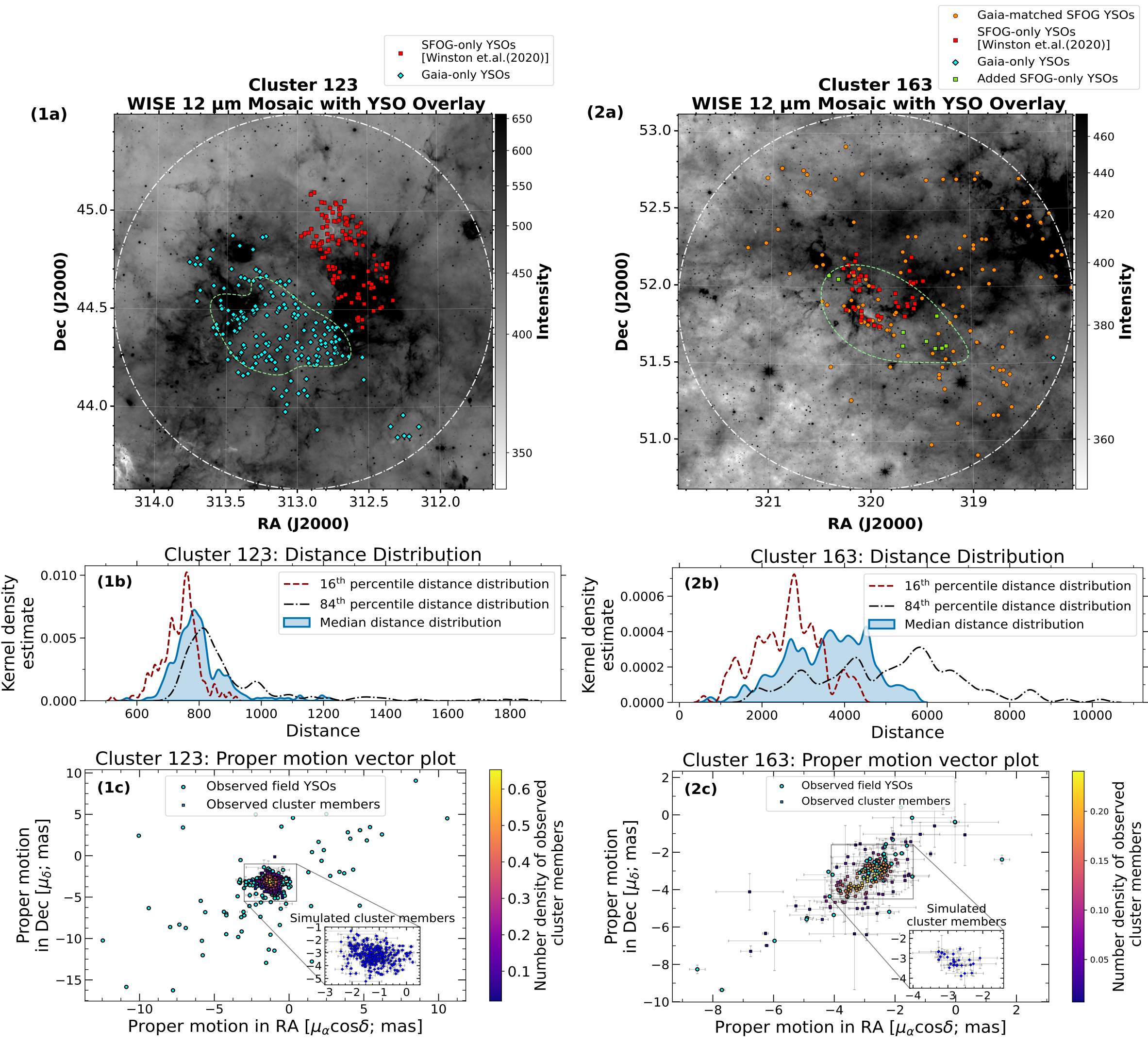} 
		\caption{Spatial and proper motion distribution of cluster members for [1] Cluster 123 and [2] Cluster 163. \textbf{(a)} The subplots display YSO cluster members overlaid on a WISE $12\mu$m inverse grayscale mosaic in the RA-Dec plane. The white dash-dotted circle indicates the region analyzed for cluster membership. Orange squares indicate Gaia-matched SFOG YSOs, while cyan diamonds represent Gaia-only cluster members from DR3. Red squares mark SFOG-only YSOs from \citet{Winston2020}. Bright green squares denote SFOG-only YSOs within the 30th- and 70th-percentile YSO density contours (green dashed curves) for Clusters 123 and 163, respectively, which are also included as cluster members. The YSOs, used to create the contour, are identified cluster members from the HDBSCAN-MC and GOLDFIST simulation and therefore are more evolved Gaia-only or Gaia-matched SFOG YSOs. \textbf{(b)} The ICRS distance distribution of cluster members is illustrated, with the blue-filled plot representing the median distance and the red dashed and black dash-dotted lines indicating the 16th and 84th percentiles, respectively.
        \textbf{(c)} Proper motion vector plot comparing observed and simulated cluster members. The color scale represents the number density of observed members (squares) selected using the Monte Carlo threshold ($\mathcal{MC}_{thr}$). Observed field YSOs are shown as cyan circles. The inset highlights simulated cluster members, while gray error bars, displayed with reduced opacity, indicate proper motion uncertainties. This subplot compares the proper motions of simulated and observed cluster members.}
		\label{fig: cluster_results}
\end{figure*}

The kinematic distance estimate for Cluster 163 is 9900 pc. The farthest cluster analyzed using Gaia data in the literature is up to 2500 pc by \cite{Zhang2023}. Limited efforts in this direction can be attributed to large distance uncertainties at these distance ranges in previous and latest Gaia releases. Therefore this work marks the very first endeavor to analyze distant outer Galaxy clusters using relatively untapped data in Gaia DR3. Figure \ref{fig: cluster_results}(2a) shows the spatial distribution of cluster members in the RA-Dec plane, with YSOs overlaid on the WISE 12 $\mu$m mosaic. Notably, only one Gaia-only YSO is detected as a cluster member. The SFOG-only YSOs, identified by \citet{Winston2020}, exhibit a strong association with the molecular cloud structure, similar to Cluster 123 and Cluster 257. This cluster appears to be more evolved, as most members are either Gaia-matched SFOG or SFOG-only (IR) YSOs.

Considering the sparse YSO distribution, we constructed a more restrictive $70^{th}$ percentile YSO density contour in the RA-Dec plane for the identification of SFOG-only cluster members. The members identified by HDBSCAN-MC and GOLDFIST simulations were used for the contour construction. Figure \ref{fig: cluster_results}(2b) presents the ICRS distance distribution for the identified Gaia-only and Gaia-matched SFOG YSOs. The median $50^{th}$ percentile distance is 3656 pc $\pm$ 1230 pc, which closely matches the more accurate simulation-predicted distance of 3531 pc $\pm$ 444 pc, as summarized in Table \ref{tab: cluster_statistics}.

However, it is important to note that both the Gaia median distance estimate and the simulation-predicted distance deviate significantly from the WISE \ion{H}{2} kinematic estimate. This discrepancy may arise because the observed cluster lies in the foreground, with background cluster stars undetected in Gaia bands due to high extinction. Alternatively, the observed members may simply be foreground field stars unrelated to any YSO cluster. The low membership fraction in the GOLDFIST simulation ($\Phi_{mem}$ = 0.14 $\pm$ 0.004) further suggests that very few YSOs are associated with the cluster if it exists at all. A more detailed analysis of such cases will require deeper IR surveys, similar to Gaia, capable of mapping the Galaxy in full six-dimensional phase space. 

\begin{figure*}[t]
    \centering
    \includegraphics[width=\linewidth]{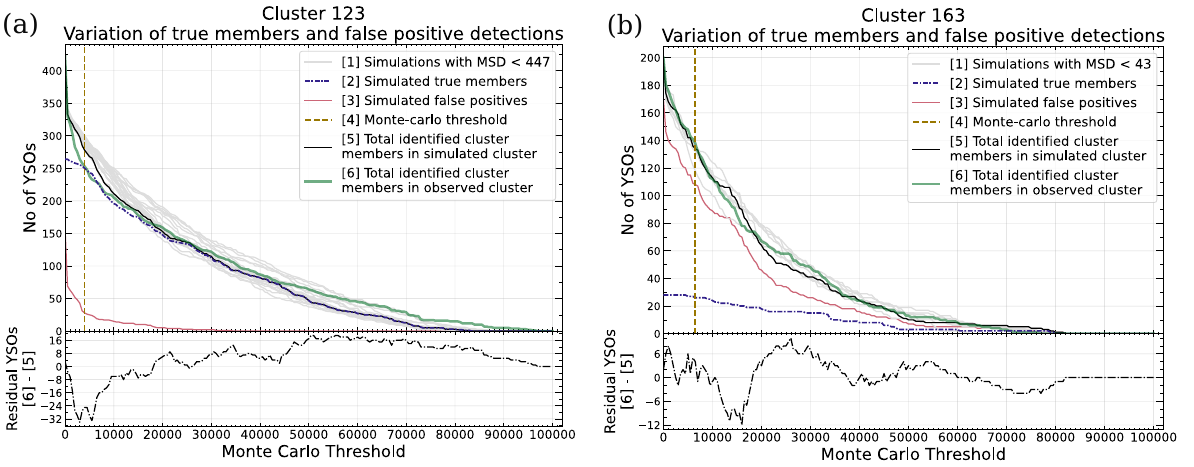} 
    \caption{The Monte Carlo spectra for field and member YSOs in (a) Cluster 123 and (b) Cluster 163. The kinematic distance estimate for Cluster 123 is unknown, with the simulation predicting a distance of 740 pc $\pm$ 6 pc. For Cluster 163, the kinematic distance estimate is 9900 pc, while the simulation predicts a distance of 3531 pc $\pm$ 444 pc. The green curve illustrates the decrease in observed total cluster members with increasing threshold. The black curve represents the identified total cluster members for the simulated cluster. The blue dash-dotted line and red curve depict the decline in simulated true members and false positives, respectively, with increasing threshold. The difference between identified total cluster members for the observed and simulated clusters is shown by the black dash-dotted line in the residual plot. The faded grey lines represent the total identified cluster members for simulated clusters generated using an alternative set of parameters, with mean squared deviation (MSD) within three times the MSD of the best-fit black curve from the green plot. The yellow vertical dash-dotted line represents the final determined Monte-Carlo threshold ($\mathcal{MC}_{thr}$), where the simulated true members comprise 95\% of the initial count. $\mathcal{MC}_{thr}$ is 4000 and 6500 for Clusters 123 and 163, respectively.}\label{fig: cummulative_Cluster_123_163}
\end{figure*}

\section{Summary and Conclusion} \label{sec:summary}

In this work, we describe a novel algorithm, HDBSCAN-MC, capable of n-dimensional unsupervised cluster membership assignment in the presence of significant astrometric uncertainties. We have demonstrated its use by identifying cluster members in outer Galaxy YSO clusters.  

We matched SFOG YSOs with Gaia DR3 and compiled a Gaia-SFOG dataset consisting of 30,738 outer Galaxy YSOs for further analysis. Additionally, we developed HDBSCAN-MC and parallelized it on the SI/HPC to perform 3D Monte Carlo simulation with 5D $\text{HDBSCAN}^{*}$ clustering. HDBSCAN-MC required Monte Carlo threshold. This was determined with cluster simulations which utilized multidimensional global genetic optimization enabling robust membership determination. The developed GOLDFIST simulations provided insights into the statistics of field YSOs and cluster members. A new distance estimate, based on simulation, was compared to kinematic and Gaia median distance estimates for three distinct clusters. As a proof of concept, we analyzed Cluster 123, Cluster 257 (W4/W5 complex) and Cluster 163. Sky positions, distance statistics, and membership records for these clusters are summarized in Table \ref{tab: cluster_statistics}. This work presents the developed methodology; a comprehensive analysis of the full set of outer Galaxy clusters will follow in a subsequent publication. Our technique is reliable and robust for cluster distance estimation under varying conditions. HDBSCAN-MC, in conjunction with the GOLDFIST simulation, provides a robust framework for systematic cluster property analysis and membership determination, making it a valuable tool for future Galactic surveys.

\section{Data Availability}
The Zenodo record \citet{patel2025} and the GitHub repository \href{https://github.com/vpatel2000/HDBMC-GOLDFIST-Sim.git}{HDBMC-GOLDFIST-Sim} hosts the Python codebase developed in this work along with the Gaia-SFOG cross-match dataset. Table \ref{tab: cluster_tables} provides the cluster tables for the analyzed clusters.

\section{Acknowledgments}
We thank the reviewers for their insightful suggestions, which have helped improve the quality of the publication. This work presents results from the European Space Agency (ESA) space mission Gaia. Gaia data are being processed by the Gaia Data Processing and Analysis Consortium (DPAC). Funding for the DPAC is provided by national institutions, in particular the institutions participating in the Gaia MultiLateral Agreement (MLA). This work is based in part on observations made with the Spitzer Space Telescope, which was operated by the Jet Propulsion Laboratory, California Institute of Technology under a contract with the National Aeronautics and Space Administration. This publication also makes use of data products from the Two Micron All Sky Survey, which is a joint project of the University of Massachusetts and the Infrared Processing and Analysis Center/California Institute of Technology, funded by the National Aeronautics and Space Administration and the National Science Foundation. This research has made use of the VizieR catalog access tool, CDS,  Strasbourg, France (DOI : 10.26093/cds/vizier). The original description of the VizieR service was published in 2000, A\&AS 143, 23. The research also made use of the cross-match service provided by CDS, Strasbourg. The computations in this work were primarily conducted on the Smithsonian High Performance Cluster (SI/HPC), Smithsonian Institution, \url{https://doi.org/10.25572/SIHPC}. We also acknowledge the usage of the high-performance computing Virgo cluster facility at the Indian Institute of Space science and Technology. Additionally, we acknowledge the use of ChatGPT (OpenAI, 2020; \url{https://openai.com/blog/chatgpt}) for paraphrasing some text and coding support. We extend our heartfelt thanks to Dr. Jagadheep D and Dr. Samir Mandal for their valuable feedback which has significantly contributed to the improvement of this work.

\vspace{5mm}
\facilities{Gaia, Spitzer (IRAC), CTIO: 1.3m, WISE, Smithsonian Institution Hydra Cluster, IIST Virgo Cluster}.

\software{Astropy \citep{astropy:2013, astropy:2018, astropy:2022}, Matplotlib \citep{Hunter:2007},
          Seaborn \citep{seaborn},
          Montage \citep{berriman2008montage},  
          $\text{HDBSCAN}^{*}$ \citep{McInnes2017},
          TOPCAT \citep{taylor2005topcat},
          Inkscape \citep{Inkscape}}.


\appendix

\section{SFOG-Gaia cross-match}\label{app: sfog-gaia}

\subsection{Definition of Galactocentric Frame}\label{app: sfog-gaia_galcen}
The Galactocentric frame is defined as follows: The Sun's position is along the x-axis, which points towards the Galactic center (l, b)=(0\degree, 0\degree). The y-axis points towards the Galactic longitude, l=90\degree, and the z-axis points towards the North Galactic Pole (b=90\degree). The Galactocentric distance of the Sun is 8.178 kpc \citep{abuter2019geometric}, with a height of 20.8 pc above the Galactic midplane. The Galactic center's ICRS coordinates are ($\alpha$, $\delta$) = ($266.4051^\circ$, $-28.936175^\circ$), and the solar motion relative to the Galactic center is (12.9, 245.6, 7.78) km/s, based on the updated default values introduced in \texttt{astropy} version 4.0.

\subsection{Chance alignment detection with photometry}
\begin{figure*}[h]
    \centering
    \includegraphics[width=0.9\linewidth]{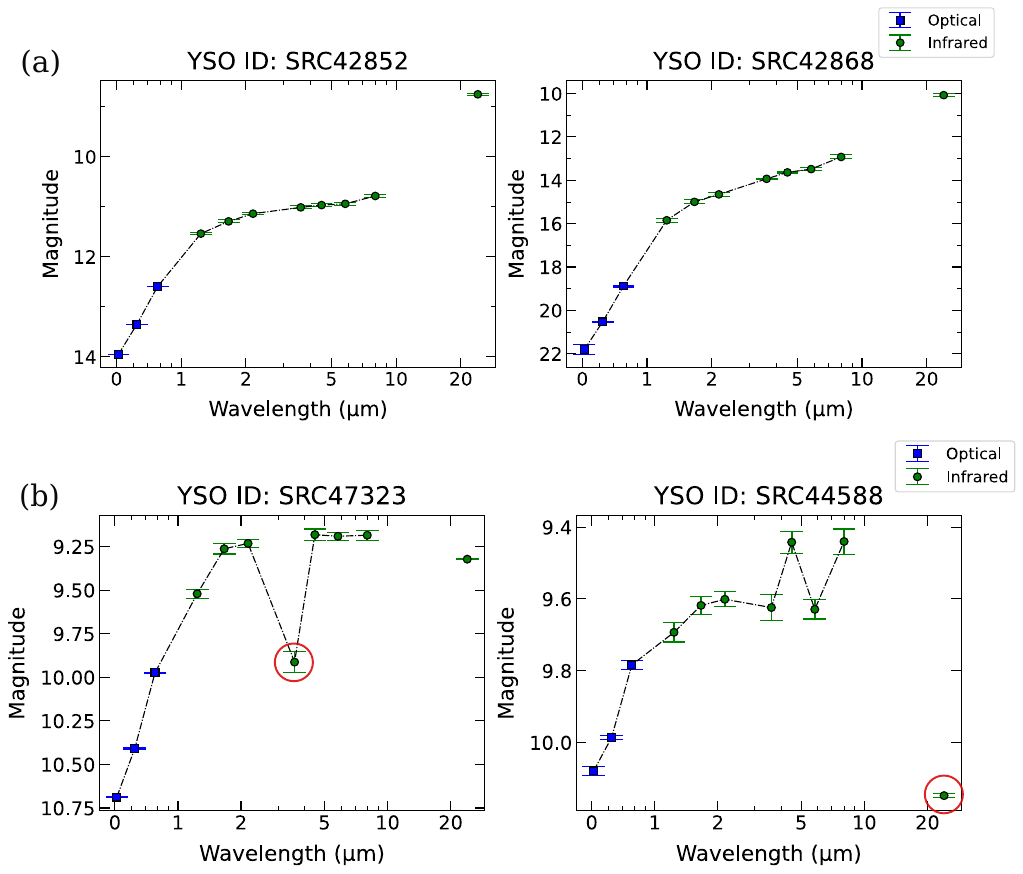} 
    \caption{Stellar spectral energy distributions (SEDs) with magnitude on the vertical axis, illustrating characteristic trends in infrared and optical magnitudes for (a) correct and (b) false SFOG-Gaia associations. The horizontal axis, plotted on a logarithmic scale, represents the wavelengths of photometric filters in micrometers ($\mu$m). Blue data points represent Gaia DR3 optical bands ($G_{BP}$, G, and $G_{RP}$). Green points correspond to infrared bands, including 2MASS J, H, Ks, IRAC bands 1 to 4, WISE W3 and W4, and MIPS $24\ \mu$m. In panel (b), the red-circled points highlight deviations from the expected YSO magnitude trend, indicating a probable chance alignment between SFOG and Gaia DR3 (see Equation \ref{eq: mag_criteria}). This figure demonstrates the photometric method used to identify 94 one-to-one chance alignments.}
		\label{fig: mag_issue_chance_good}
\end{figure*}

The one-to-one and many-to-one chance alignments between the Gaia-SFOG cross-match are resolved with the photometry-based approach described in subsection \ref{subsec: crossmatch_DR3}. Figure \ref{fig: mag_issue_chance_good} illustrates characteristic
trends in infrared and optical magnitudes for (a) correct and (b) false SFOG-Gaia associations based on Equation \ref{eq: mag_criteria}.

\section{Astrometric Uncertainties and Optimization Details for  GOLDFIST Framework\label{app: goldfist}}

The distance and proper motion uncertainties are incorporated by assigning observed data to the simulated YSOs. Details of this procedure are provided in subsections \ref{app: goldfist_dist} and \ref{app: goldfist_prop} of this section. Subsection \ref{app: goldfist_spatial_opt} provides additional spatial optimization details.

\subsection{Simulating distance uncertainties\label{app: goldfist_dist}}
To account for errors in Gaia distance estimates, the simulated distances were perturbed by random sampling from fixed percentile bins, corresponding to the distance error distribution of each YSO. The distance error distribution for each YSO was modeled using an Inverse Weibull distribution. The percentile bins used for sampling were ($80^{th}$, $77^{th}$), ($68^{th}$, $65^{th}$), ($30^{th}$, $27^{th}$), ($15^{th}$, $12^{th}$), and ($5^{th}$, $3^{rd}$). The actual errors from the observed data were then assigned to the simulated YSOs using a modified Jonker-Volgenant algorithm, with the cost matrix based on the synthetic distance $\rho_{\rm err_{1}}$ (Equation \ref{eq:rho_1}). The entire Gaia-SFOG dataset, comprising 30,738 YSOs, was used for this assignment. This perturbation introduced the observed stretch in the cluster and field YSO distribution along the ICRS distance axis.

In Equation \ref{eq:rho_1}, $i=\{1,2,3\}$ refers to the distance (simulated and observed median ICRS YSO distances), Gaia $G$-band magnitude, and $B_{P}-R_{P}$ color, respectively. The observables were normalized with Median Absolute Deviation (MAD) to bring them on the same scale.

\begin{equation}
    \rho_{\rm err_{1}} = \sum_{i=1}^{3} 
    \sqrt{\bigg[\frac{\Delta i}{MAD(i)}\bigg]^2}\label{eq:rho_1}
\end{equation}

Implementing the observed stretch in the simulated YSO cluster along the distance axis altered the ideal YSO positions in the simulation, rendering the initially assigned distance errors invalid for the updated locations. Therefore, final distance errors were reassigned following the same technique. This time, the modified Jonker-Volgenant algorithm was applied using a cost matrix based on the synthetic distance $\rho_{\rm err_{2}}$ (Equation \ref{eq:rho_2}). For this assignment, only the observed data within the RA-Dec space of the analyzed cluster was used.
\footnote{Let $d_{16}$, $d_{50}$, and $d_{84}$ denote $16^{th}$, $50^{th}$ and $84^{th}$ distance error percentiles respectively for the observed YSOs. Assignment of observed distance errors to the simulated data involves adding $d_{84} - d_{50}$ and subtracting $d_{50} - d_{16}$ from the simulated median distance to obtain the simulated upper ($84^{\text{th}}$ percentile) and lower ($16^{\text{th}}$ percentile) uncertainty limits. If subtracting $d_{50} - d_{16}$ from the simulated median distance results in a negative value, the assignment for such simulated YSO is taken from the entire Gaia-SFOG dataset.}

In Equation \ref{eq:rho_2}, $j=\{1,2,3\}$ refers to the distance (perturbed simulated and observed median ICRS YSO distances), Galactic longitude ($l$), and Galactic latitude ($b$), respectively.

\begin{equation}
    \rho_{\rm err_{2}} = \sum_{j=1}^{3} 
    \sqrt{\bigg[\frac{\Delta j}{MAD(j)}\bigg]^2} \label{eq:rho_2}
\end{equation}

\subsection{Simulating proper motion uncertainties}\label{app: goldfist_prop}
The following approach was adopted to model proper motion uncertainties for simulated YSOs. The observed proper motion errors of YSOs from the analysis region were assigned to the simulated cluster YSOs based on their closest match in terms of mean proper motion values. The modified Jonker-Volgenant algorithm efficiently executed this process, utilizing a cost matrix based on the synthetic distance $\rho_{\rm pm}$ (Equation \ref{eq: rho_pm}).

\begin{equation}
    \rho_{\rm pm} = \sqrt{\big[\Delta \rm \mu_{\alpha}cos\delta\big]^2 + \big[\Delta \mu_{\delta}\big]^2}
    \label{eq: rho_pm}
\end{equation}

During this assignment, in each optimization iteration, the proper motion errors of the confirmed field YSOs which corresponded to the predicted cluster distance range [\textbf{$\mathcal{D}_{center}$} - $r_t$, \textbf{$\mathcal{D}_{center}$} + $r_t$] were not assigned to the simulated cluster members. The proper motion and errors of the remaining observed YSOs (i.e. observed YSOs not assigned to simulated cluster members by cost matrix based on Equation \ref{eq: rho_pm}) were assigned to simulated field YSOs. The process utilized a cost matrix based on the differences between observed median and simulated perturbed YSO distances. The same modified Jonker-Volgenant assignment as described previously was used for this purpose.

\subsection{Additional Details for Spatial Optimization\label{app: goldfist_spatial_opt}}

All six coordinate sets—namely, the observed Galactocentric coordinates (X, Y, Z), representing the spatial positions of YSOs in the analyzed region, and the simulated Galactocentric coordinates (x, y, z)—share the same length, denoted by 
$L$, which corresponds to the total number of YSOs in the region. Subsequently, histograms are constructed for each set of coordinates, utilizing a number of bins equal to $L/5$. This choice strikes a balance, as fewer bins could result in information loss, while more bins might overly emphasize individual YSO coordinates rather than the overall distribution.

The frequency of YSOs within each bin is normalized by the total number of YSOs ($L$), to form the probability distribution. For instance, for the set X, the probability distribution, $\mathds{P}(\rm X)$ is calculated as $\text{hist}(\text{X})/L$, where $\text{hist}(\rm X)$ represents histogram frequencies for X coordinates. This normalization process is replicated for sets Y, Z, x, y, and z. However, a small constant ($10^{-9}$) is added to the x, y, and z probability values. This addition ensures that zero probability is not encountered in the objective function (F, Equation \ref{eq:f}), which utilizes Kullback-Leibler (KL) divergence.

The objective function $F$ (Equation \ref{eq:f}), subject to the constraints (Equation \ref{eq:f_constraints}) is minimized using Differential Evolution. In this process, the \textbf{$\mathcal{D}_{center}$} parameter is constrained within a 20\% kinematic distance error range. However, for clusters lacking a known kinematic distance estimate, the distance bounds are defined as (D - m$\sigma$, D + n$\sigma$), where m and n are integers. Here D and $\sigma$ refer to the median Gaia distance estimate and the average symmetrical distance uncertainty of member YSOs obtained by setting $\mathcal{MC}_{thr} = 50,000$. The membership fraction (\textbf{$\Phi_{mem}$}) typically ranges between 0.5 and 0.99 for most clusters, as a significant portion of YSOs in the field are expected to be cluster members. However, for distant or sparsely populated RA-Dec fields, \textbf{$\Phi_{mem}$} may be adjusted to (0.1, 0.99), allowing exploration of simulations where noise dominates over true cluster members.
The cluster concentration parameter, \textbf{$\xi_{c}$} ranges from 0.01 to 0.99. Table \ref{tab:parameter_table} gives default bounds for upper and lower limits of the field YSO distribution along the ICRS distance axis i.e. \textbf{$F_{\mathcal{L}}$} and \textbf{$F_{\mathcal{U}}$} parameters.

\section{Optimization runs and parameter values for the analyzed clusters\label{app: DE_tables}}

The mean squared deviation (MSD), calculated after both stages of optimization, quantifies the deviation of the simulated Monte Carlo spectrum from the observed spectrum (Figures \ref{fig: mc_spectra}, \ref{fig: cummulative_Cluster_123_163}(a), and \ref{fig: cummulative_Cluster_123_163}(b)). Table \ref{tab: params} presents the optimized and the model-predicted weighted parameter values for parameter sets that satisfy $\text{MSD} < 3 \times \text{MSD}_{\text{min}}$ for each cluster. These were chosen from the top 30 spatial optimization matches (subsection \ref{subsec: spatial_optimization}) that were selected for proper motion optimization (subsection \ref{subsec: pm_optimization}) from a total of 60 performed optimizations. 

\begin{deluxetable}{cccccccc}[h!]
\tablecaption{Optimization runs and parameter values for analyzed clusters \label{tab: params}}
\tablewidth{\textwidth}
\tabletypesize{\tiny}
\tablehead{
\multicolumn{8}{c}{\rule{0pt}{4.8ex} \footnotesize Cluster 123} \\  
\hline
\colhead{\textbf{Run \#}} & \colhead{\textbf{$\mathcal{D}_{center}$}} & \colhead{\textbf{$\Phi_{mem}$}} & \colhead{\textbf{$F_{\mathcal{L}}$}} & \colhead{\textbf{$F_{\mathcal{U}}$}} & 
\colhead{\textbf{$\xi_{c}$}} & \colhead{\textbf{$\Lambda_{pm}$}} & \colhead{\textbf{MSD}}
}
\startdata
27 & 744 & 0.61 & 502 & 1615 & 0.619 & 2.881 & 148.975 \\
6  & 743 & 0.61 & 580 & 1492 & 0.588 & 2.877 & 184.503 \\
9  & 743 & 0.61 & 568 & 1643 & 0.446 & 2.888 & 193.249 \\
24 & 743 & 0.61 & 602 & 1628 & 0.612 & 2.887 & 209.527 \\
4  & 744 & 0.61 & 521 & 1503 & 0.448 & 2.903 & 224.463 \\
15 & 743 & 0.61 & 626 & 1527 & 0.381 & 2.876 & 234.801 \\
7  & 739 & 0.59 & 630 & 1691 & 0.828 & 2.831 & 249.323 \\
17 & 743 & 0.61 & 500 & 1532 & 0.441 & 2.880 & 267.448 \\
19 & 741 & 0.61 & 552 & 1509 & 0.579 & 2.849 & 309.443 \\
16 & 742 & 0.61 & 502 & 1574 & 0.431 & 2.882 & 318.294 \\
3  & 725 & 0.51 & 568 & 1625 & 0.763 & 3.165 & 326.617 \\
12 & 739 & 0.59 & 595 & 1706 & 0.813 & 3.089 & 334.632 \\
13 & 741 & 0.61 & 644 & 1517 & 0.434 & 2.873 & 339.587 \\
26 & 741 & 0.61 & 496 & 1525 & 0.400 & 2.917 & 361.368 \\
5  & 741 & 0.61 & 518 & 1509 & 0.435 & 2.858 & 370.702 \\
28 & 746 & 0.64 & 545 & 1701 & 0.860 & 3.352 & 378.304 \\
23 & 727 & 0.51 & 492 & 1631 & 0.535 & 3.359 & 410.876 \\
29 & 729 & 0.51 & 525 & 1424 & 0.461 & 3.196 & 412.244 \\
22 & 728 & 0.51 & 641 & 1556 & 0.386 & 3.360 & 444.343 \\
\hline
\textbf{Average} & 740 & 0.59 & 558 & 1577 & 0.554 & 2.962 & \\
\textbf{Unbiased SD} & 6 & 0.04 & 52 & 77 & 0.152 & 0.168 & \\
\hline
\multicolumn{8}{c}{\rule{0pt}{3.5ex} \footnotesize Cluster 257} \\ 
\hline
\colhead{\textbf{Run \#}} & \colhead{\textbf{$\mathcal{D}_{center}$}} & \colhead{\textbf{$\Phi_{mem}$}} & \colhead{\textbf{$F_{\mathcal{L}}$}} & \colhead{\textbf{$F_{\mathcal{U}}$}} & 
\colhead{\textbf{$\xi_{c}$}} & \colhead{\textbf{$\Lambda_{pm}$}} & \colhead{\textbf{MSD}} \\
\hline
5  & 2379 & 0.58 & 561  & 3878 & 0.579 & 5.689 & 242.294 \\
14 & 1963 & 0.59 & 715  & 3582 & 0.688 & 4.158 & 268.134 \\
25 & 1994 & 0.59 & 1040 & 3531 & 0.532 & 4.094 & 333.503 \\
12 & 2265 & 0.61 & 1089 & 3706 & 0.563 & 4.825 & 426.552 \\
24 & 1925 & 0.57 & 971  & 3365 & 0.478 & 3.773 & 430.741 \\
17 & 2276 & 0.50 & 710  & 3601 & 0.558 & 4.391 & 464.080 \\
18 & 2331 & 0.60 & 903  & 4455 & 0.966 & 4.583 & 477.353 \\
13 & 2215 & 0.54 & 634  & 4028 & 0.621 & 4.700 & 483.383 \\
6  & 2269 & 0.52 & 458  & 4245 & 0.763 & 4.384 & 519.876 \\
23 & 1931 & 0.58 & 627  & 3764 & 0.628 & 3.672 & 527.333 \\
11 & 2306 & 0.65 & 1097 & 3463 & 0.580 & 5.104 & 638.995 \\
27 & 2248 & 0.54 & 1003 & 3898 & 0.258 & 4.312 & 648.010 \\
8  & 1967 & 0.56 & 898  & 3404 & 0.864 & 3.650 & 677.776 \\
\hline
\textbf{Average} & 2158 & 0.57 & 804 & 3757 & 0.619 & 4.482 & \\
\textbf{Unbiased SD} & 180 & 0.04 & 216 & 314 & 0.158 & 0.641 & \\
\hline
\multicolumn{8}{c}{\rule{0pt}{3.5ex} \footnotesize Cluster 163} \\ 
\hline
\colhead{\textbf{Run \#}} & \colhead{\textbf{$\mathcal{D}_{center}$}} & \colhead{\textbf{$\Phi_{mem}$}} & \colhead{\textbf{$F_{\mathcal{L}}$}} & \colhead{\textbf{$F_{\mathcal{U}}$}} & 
\colhead{\textbf{$\xi_{c}$}} & \colhead{\textbf{$\Lambda_{pm}$}} & \colhead{\textbf{MSD}} \\
\hline
3  & 4006 & 0.14 & 1680  & 6731 & 0.050 & 8.368 & 14.244 \\
2  & 2988 & 0.14 & 2140  & 6754 & 0.862 & 6.089 & 16.388 \\
17 & 3692 & 0.14 & 1679  & 6744 & 0.884 & 7.740 & 20.134 \\
4  & 3116 & 0.14 & 2018  & 6669 & 0.535 & 6.048 & 21.259 \\
26 & 3714 & 0.13 & 2107  & 6913 & 0.753 & 5.844 & 27.313 \\
24 & 3990 & 0.13 & 1800  & 6911 & 0.366 & 7.787 & 28.567 \\
5  & 3192 & 0.14 & 1986  & 7082 & 0.904 & 6.746 & 29.403 \\
\hline
\textbf{Average} & 3531 & 0.14 & 1902 & 6803 & 0.589 & 7.027 & \\
\textbf{Unbiased SD} & 444 & 0.004 & 208 & 133 & 0.355 & 1.072 & \\
\enddata
\tablecomments{
The topmost parameter set for each analyzed region corresponds to the cluster simulation with the minimum MSD between the observed and simulated Monte Carlo spectrum. 'Run \#' represents the ranking of the parameter set among the top 30 spatial optimization matches that were selected for proper motion optimization from 60 performed total optimizations. All listed parameter sets for each cluster satisfy $\text{MSD} < 3 \times \text{MSD}_{\text{min}}$ and are ordered by ascending MSD. 'Average' and 'Unbiased SD' represent the weighted mean and weighted unbiased standard deviation, respectively.}
\end{deluxetable}

\section{Cluster tables\label{app: cluster_tables}}

The cluster tables for Clusters 123, 163, and W4/W5 complex along with their metadata, are presented in Table \ref{tab: cluster_tables}.

\begin{deluxetable}{ccc}[h!]
\tablecaption{Metadata for Cluster tables\label{tab: cluster_tables}}
\tablewidth{0pt}
\tabletypesize{\tiny}
\tablehead{
\colhead{Column Number} & \colhead{Column Name} & \colhead{Description}
}
\startdata
1  & ID                     & Unique source identifier \\
2  & IRname                & Infrared source name (e.g., IRAS) \\
3  & 2Mname                & 2MASS ID \\
4  & Cluster       & Cluster designation from W2020 \\
5  & Class         & YSO classification from W2020 \\
6  & RAdeg          & Right ascension (J2000) [deg]\\
7  & RAerr   & Right ascension uncertainty (J2000)[mas]\\
8  & DEdeg         & Declination (J2000) [deg]\\
9  & DEerr  & Declination uncertainty (J2000) [mas]\\
10 & Jmag                      & J-band magnitude (Vega)\\
11 & e\_Jmag                 & J-band magnitude uncertainty (Vega)\\
12 & fJ          & J-band flux (mJy)\\
13 & dfJ         & J-band flux uncertainty (mJy)\\
14 & Hmag                      & H-band magnitude (Vega)\\
15 & e\_Hmag                 & H-band magnitude uncertainty (Vega)\\
16 & fH                   & H-band flux (mJy)\\
17 & e\_fH                  & H-band flux uncertainty (mJy)\\
18 & Ksmag                     & Ks-band magnitude (Vega)\\
19 & e\_Ksmag                & Ks-band magnitude uncertainty (Vega)\\
20 & fKs                   & Ks-band flux (mJy)\\
21 & e\_Ks                  & Ks-band flux uncertainty (mJy)\\
22 & 3.6mag                & IRAC band 1 magnitude (Vega)\\
23 & e\_3.6mag                 & IRAC band 1 magnitude uncertainty (Vega)\\
24 & f3.6                    & IRAC band 1 flux (mJy)\\
25 & e\_3.6                   & IRAC band 1 flux uncertainty (mJy)\\
26 & 4.5mag                & IRAC band 2 magnitude (Vega)\\
27 & e\_4.5mag                 & IRAC band 2 magnitude uncertainty (Vega)\\
28 & f4.5                    & IRAC band 2 flux (mJy)\\
29 & e\_f4.5                   & IRAC band 2 flux uncertainty (mJy)\\
30 & W3mag                 & WISE W3 magnitude (Vega)\\
31 & e\_W3mag              & WISE W3 magnitude uncertainty (Vega) \\
32 & W4mag                & WISE W4 magnitude (Vega)\\
33 & e\_W4mag              & WISE W4 magnitude uncertainty (Vega)\\
34 & GaiaDR3            & Gaia DR3 designation \\
35 & Source\_ID             & Gaia DR3 source identifier \\
36 & plx               & Gaia DR3 parallax [mas]\\
37 & e\_plx         & Gaia DR3 parallax uncertainty [mas]\\
38 & plxOverErr             & Parallax over error [mas]\\
39 & pmRA                   & Proper motion in RA [mas]\\
40 & e\_pmRA                   & Uncertainty in pmRA\\
41 & pmDEC                  & Proper motion in Dec [mas]\\
42 & e\_pmDEC               & Uncertainty in pmDEC\\
43 & RUWE & Renormalized unit weight error\\
44 & Gmag     & Gaia G-band mean magnitude (Vega)\\
45 & e\_Gmag                & Gaia G-band magnitude uncertainty(Vega)\\
46 & BPmag     & Gaia BP mean magnitude (Vega)\\
47 & e\_BPmag                & Gaia BP magnitude uncertainty(Vega)\\
48 & RPmag     & Gaia RP mean magnitude (Vega)\\
49 & e\_RPmag                & Gaia RP magnitude uncertainty(Vega)\\
50 & Bp-Rp                 & GaiaBP-RP color (Vega)\\
51 & Bp-G                & Gaia BP-G color (Vega)\\
52 & G-Rp                 & Gaia G-RP color (Vega)\\
53 & dist            & ICRS source distance [pc]\\
54 & b\_dist        & Lower bound on dist [pc] \\
55 & B\_dist        & Upper bound on dist [pc] \\
56 & X                      & Galactocentric X-coordinate [pc]\\
57 & Y                      & Galactocentric Y-coordinate [pc]\\
58 & Z                      & Galactocentric Z-coordinate [pc]\\
59 & mem\_flag              & Membership flag for YSO cluster \\
60 & Prob              & Membership probability \\
61 & e\_Prob         & Uncertainty in Prob \\
\enddata
\tablecomments{W2020 refers to \citet{Winston2020}. Cluster members are marked with \texttt{mem\_flag = 1}, while \texttt{mem\_flag = 0} indicates field YSOs. \texttt{mem\_flag = 2} corresponds to SFOG-only YSOs identified as cluster members in \citet{Winston2020}, and \texttt{mem\_flag = 3} denotes additional SFOG-only cluster members identified in this work based on the $30^{th}$ ($70^{th}$ for Cluster 163) percentile YSO density contour (see Section \ref{sec: poc}). YSOs not seen by Gaia have \texttt{source\_id = -1}. Complete tables for Cluster 123, the W4/W5 complex (Cluster 257), and Cluster 163 are available as a downloadable machine-readable file.}

\end{deluxetable}

\FloatBarrier

\bibliography{AAS_I}{}

\begin{thebibliography}{}
\expandafter\ifx\csname natexlab\endcsname\relax\def\natexlab#1{#1}\fi
\providecommand{\url}[1]{\href{#1}{#1}}
\providecommand{\dodoi}[1]{doi:~\href{http://doi.org/#1}{\nolinkurl{#1}}}
\providecommand{\doeprint}[1]{\href{http://ascl.net/#1}{\nolinkurl{http://ascl.net/#1}}}
\providecommand{\doarXiv}[1]{\href{https://arxiv.org/abs/#1}{\nolinkurl{https://arxiv.org/abs/#1}}}

\bibitem[{R. Abuter {et~al.}(2019)Abuter, Amorim, Baub{\"o}ck, Berger, Bonnet, Brandner, Cl{\'e}net, Du~Foresto, De~Zeeuw, Dexter, {et~al.}}]{abuter2019geometric}
Abuter, R., Amorim, A., Baub{\"o}ck, M., {et~al.} 2019, \bibinfo{title}{A geometric distance measurement to the Galactic center black hole with 0.3\% uncertainty,} Astronomy \& Astrophysics, 625, L10, \dodoi{https://doi.org/10.1051/0004-6361/201935656}

\bibitem[{R. Andrae {et~al.}(2023)Andrae, Fouesneau, Sordo, Bailer-Jones, Dharmawardena, Rybizki, Angeli, Lindstrøm, Marshall, Drimmel, Korn, Soubiran, Brouillet, Casamiquela, Rix, Aramburu, Álvarez, Bakker, Bellas-Velidis, Bijaoui, Brugaletta, Burlacu, Carballo, Chaoul, Chiavassa, Contursi, Cooper, Creevey, Dafonte, Dapergolas, Laverny, Delchambre, Demouchy, Edvardsson, Frémat, Garabato, García-Lario, García-Torres, Gavel, Gomez, González-Santamaría, Hatzidimitriou, Heiter, Piccolo, Kontizas, Kordopatis, Lanzafame, Lebreton, Licata, Livanou, Lobel, Lorca, Romeo, Manteiga, Marocco, Mary, Nicolas, Ordenovic, Pailler, Palicio, Pallas-Quintela, Panem, Pichon, Poggio, Recio-Blanco, Riclet, Robin, Santoveña, Sarro, Schultheis, Segol, Silvelo, Slezak, Smart, Süveges, Thévenin, Elipe, Ulla, Utrilla, Vallenari, Dillen, Zhao, \& Zorec}]{Andrae2023}
Andrae, R., Fouesneau, M., Sordo, R., {et~al.} 2023, \bibinfo{title}{Gaia Data Release 3: Analysis of the Gaia BP/RP spectra using the General Stellar Parameterizer from Photometry,} Astronomy and Astrophysics, 674, \dodoi{10.1051/0004-6361/202243462}

\bibitem[{ {Astropy Collaboration} {et~al.}(2013){Astropy Collaboration}, {Robitaille}, {Tollerud}, {Greenfield}, {Droettboom}, {Bray}, {Aldcroft}, {Davis}, {Ginsburg}, {Price-Whelan}, {Kerzendorf}, {Conley}, {Crighton}, {Barbary}, {Muna}, {Ferguson}, {Grollier}, {Parikh}, {Nair}, {Unther}, {Deil}, {Woillez}, {Conseil}, {Kramer}, {Turner}, {Singer}, {Fox}, {Weaver}, {Zabalza}, {Edwards}, {Azalee Bostroem}, {Burke}, {Casey}, {Crawford}, {Dencheva}, {Ely}, {Jenness}, {Labrie}, {Lim}, {Pierfederici}, {Pontzen}, {Ptak}, {Refsdal}, {Servillat}, \& {Streicher}}]{astropy:2013}
{Astropy Collaboration}, {Robitaille}, T.~P., {Tollerud}, E.~J., {et~al.} 2013, \bibinfo{title}{{Astropy: A community Python package for astronomy},} \aap, 558, A33, \dodoi{10.1051/0004-6361/201322068}

\bibitem[{ {Astropy Collaboration} {et~al.}(2018){Astropy Collaboration}, {Price-Whelan}, {Sip{\H{o}}cz}, {G{\"u}nther}, {Lim}, {Crawford}, {Conseil}, {Shupe}, {Craig}, {Dencheva}, {Ginsburg}, {VanderPlas}, {Bradley}, {P{\'e}rez-Su{\'a}rez}, {de Val-Borro}, {Aldcroft}, {Cruz}, {Robitaille}, {Tollerud}, {Ardelean}, {Babej}, {Bach}, {Bachetti}, {Bakanov}, {Bamford}, {Barentsen}, {Barmby}, {Baumbach}, {Berry}, {Biscani}, {Boquien}, {Bostroem}, {Bouma}, {Brammer}, {Bray}, {Breytenbach}, {Buddelmeijer}, {Burke}, {Calderone}, {Cano Rodr{\'\i}guez}, {Cara}, {Cardoso}, {Cheedella}, {Copin}, {Corrales}, {Crichton}, {D'Avella}, {Deil}, {Depagne}, {Dietrich}, {Donath}, {Droettboom}, {Earl}, {Erben}, {Fabbro}, {Ferreira}, {Finethy}, {Fox}, {Garrison}, {Gibbons}, {Goldstein}, {Gommers}, {Greco}, {Greenfield}, {Groener}, {Grollier}, {Hagen}, {Hirst}, {Homeier}, {Horton}, {Hosseinzadeh}, {Hu}, {Hunkeler}, {Ivezi{\'c}}, {Jain}, {Jenness}, {Kanarek}, {Kendrew}, {Kern}, {Kerzendorf}, {Khvalko}, {King}, {Kirkby}, {Kulkarni},
  {Kumar}, {Lee}, {Lenz}, {Littlefair}, {Ma}, {Macleod}, {Mastropietro}, {McCully}, {Montagnac}, {Morris}, {Mueller}, {Mumford}, {Muna}, {Murphy}, {Nelson}, {Nguyen}, {Ninan}, {N{\"o}the}, {Ogaz}, {Oh}, {Parejko}, {Parley}, {Pascual}, {Patil}, {Patil}, {Plunkett}, {Prochaska}, {Rastogi}, {Reddy Janga}, {Sabater}, {Sakurikar}, {Seifert}, {Sherbert}, {Sherwood-Taylor}, {Shih}, {Sick}, {Silbiger}, {Singanamalla}, {Singer}, {Sladen}, {Sooley}, {Sornarajah}, {Streicher}, {Teuben}, {Thomas}, {Tremblay}, {Turner}, {Terr{\'o}n}, {van Kerkwijk}, {de la Vega}, {Watkins}, {Weaver}, {Whitmore}, {Woillez}, {Zabalza}, \& {Astropy Contributors}}]{astropy:2018}
{Astropy Collaboration}, {Price-Whelan}, A.~M., {Sip{\H{o}}cz}, B.~M., {et~al.} 2018, \bibinfo{title}{{The Astropy Project: Building an Open-science Project and Status of the v2.0 Core Package},} \aj, 156, 123, \dodoi{10.3847/1538-3881/aabc4f}

\bibitem[{ {Astropy Collaboration} {et~al.}(2022){Astropy Collaboration}, {Price-Whelan}, {Lim}, {Earl}, {Starkman}, {Bradley}, {Shupe}, {Patil}, {Corrales}, {Brasseur}, {N{\"o}the}, {Donath}, {Tollerud}, {Morris}, {Ginsburg}, {Vaher}, {Weaver}, {Tocknell}, {Jamieson}, {van Kerkwijk}, {Robitaille}, {Merry}, {Bachetti}, {G{\"u}nther}, {Aldcroft}, {Alvarado-Montes}, {Archibald}, {B{\'o}di}, {Bapat}, {Barentsen}, {Baz{\'a}n}, {Biswas}, {Boquien}, {Burke}, {Cara}, {Cara}, {Conroy}, {Conseil}, {Craig}, {Cross}, {Cruz}, {D'Eugenio}, {Dencheva}, {Devillepoix}, {Dietrich}, {Eigenbrot}, {Erben}, {Ferreira}, {Foreman-Mackey}, {Fox}, {Freij}, {Garg}, {Geda}, {Glattly}, {Gondhalekar}, {Gordon}, {Grant}, {Greenfield}, {Groener}, {Guest}, {Gurovich}, {Handberg}, {Hart}, {Hatfield-Dodds}, {Homeier}, {Hosseinzadeh}, {Jenness}, {Jones}, {Joseph}, {Kalmbach}, {Karamehmetoglu}, {Ka{\l}uszy{\'n}ski}, {Kelley}, {Kern}, {Kerzendorf}, {Koch}, {Kulumani}, {Lee}, {Ly}, {Ma}, {MacBride}, {Maljaars}, {Muna}, {Murphy}, {Norman},
  {O'Steen}, {Oman}, {Pacifici}, {Pascual}, {Pascual-Granado}, {Patil}, {Perren}, {Pickering}, {Rastogi}, {Roulston}, {Ryan}, {Rykoff}, {Sabater}, {Sakurikar}, {Salgado}, {Sanghi}, {Saunders}, {Savchenko}, {Schwardt}, {Seifert-Eckert}, {Shih}, {Jain}, {Shukla}, {Sick}, {Simpson}, {Singanamalla}, {Singer}, {Singhal}, {Sinha}, {Sip{\H{o}}cz}, {Spitler}, {Stansby}, {Streicher}, {{\v{S}}umak}, {Swinbank}, {Taranu}, {Tewary}, {Tremblay}, {de Val-Borro}, {Van Kooten}, {Vasovi{\'c}}, {Verma}, {de Miranda Cardoso}, {Williams}, {Wilson}, {Winkel}, {Wood-Vasey}, {Xue}, {Yoachim}, {Zhang}, {Zonca}, \& {Astropy Project Contributors}}]{astropy:2022}
{Astropy Collaboration}, {Price-Whelan}, A.~M., {Lim}, P.~L., {et~al.} 2022, \bibinfo{title}{{The Astropy Project: Sustaining and Growing a Community-oriented Open-source Project and the Latest Major Release (v5.0) of the Core Package},} \apj, 935, 167, \dodoi{10.3847/1538-4357/ac7c74}

\bibitem[{C. {Babusiaux} {et~al.}(2023){Babusiaux}, {Fabricius}, {Khanna}, {Muraveva}, {Reyl{\'e}}, {Spoto}, {Vallenari}, {Luri}, {Arenou}, {{\'A}lvarez}, {Anders}, {Antoja}, {Balbinot}, {Barache}, {Bauchet}, {Bossini}, {Busonero}, {Cantat-Gaudin}, {Carrasco}, {Dafonte}, {Diakit{\'e}}, {Figueras}, {Garcia-Gutierrez}, {Garofalo}, {Helmi}, {Jim{\'e}nez-Arranz}, {Jordi}, {Kervella}, {Kostrzewa-Rutkowska}, {Leclerc}, {Licata}, {Manteiga}, {Masip}, {Mongui{\'o}}, {Ramos}, {Robichon}, {Robin}, {Romero-G{\'o}mez}, {S{\'a}ez}, {Santove{\~n}a}, {Spina}, {Torralba Elipe}, \& {Weiler}}]{gaia_2023_validation}
{Babusiaux}, C., {Fabricius}, C., {Khanna}, S., {et~al.} 2023, \bibinfo{title}{{Gaia Data Release 3. Catalogue validation},} \aap, 674, A32, \dodoi{10.1051/0004-6361/202243790}

\bibitem[{C. Bailer-Jones {et~al.}(2021)Bailer-Jones, Rybizki, Fouesneau, Demleitner, \& Andrae}]{bailer2021estimating}
Bailer-Jones, C., Rybizki, J., Fouesneau, M., Demleitner, M., \& Andrae, R. 2021, \bibinfo{title}{Estimating distances from parallaxes. V. Geometric and photogeometric distances to 1.47 billion stars in Gaia Early Data Release 3,} The Astronomical Journal, 161, 147

\bibitem[{G. {Baume} {et~al.}(2003){Baume}, {V{\'a}zquez}, {Carraro}, \& {Feinstein}}]{baume2003photometric}
{Baume}, G., {V{\'a}zquez}, R.~A., {Carraro}, G., \& {Feinstein}, A. 2003, \bibinfo{title}{{Photometric study of the young open cluster NGC 3293},} \aap, 402, 549, \dodoi{10.1051/0004-6361:20030223}

\bibitem[{G. Berriman {et~al.}(2008)Berriman, Good, Laity, \& Kong}]{berriman2008montage}
Berriman, G., Good, J., Laity, A., \& Kong, M. 2008, in Astronomical Data Analysis Software and Systems XVII, Vol. 394, 83

\bibitem[{J.~B.~G.~M. {Bloemen} {et~al.}(1984){Bloemen}, {Bennett}, {Bignami}, {Caraveo}, {Strong}, {Blitz}, {Gottwald}, {Mayer-Hasselwander}, {Hermsen}, \& {Lebrun}}]{bloemen1984radial}
{Bloemen}, J.~B.~G.~M., {Bennett}, K., {Bignami}, G.~F., {et~al.} 1984, \bibinfo{title}{{The radial distribution of galactic gamma rays. II - The distribution of cosmic-ray electrons and nuclei in the outer galaxy},} \aap, 135, 12

\bibitem[{S.~S. Boyd \& L. Vandenberghe(2004)Boyd \& Vandenberghe}]{Boyd2004}
Boyd, S.~S., \& Vandenberghe, L. 2004, Convex optimizationBoyd, S. S., \& Vandenberghe, L. (2004). Convex optimization. Optimization Methods and Software (Vol. 25). Cambridge University Press. http://doi.org/10.1080/10556781003625177, Vol.~25

\bibitem[{N. Brouillet {et~al.}(1997)Brouillet, Kaufman, Combes, Baudry, \& Bash}]{brouillet1997identification}
Brouillet, N., Kaufman, M., Combes, F., Baudry, A., \& Bash, F. 1997, \bibinfo{title}{Identification of molecular complexes in M81,} arXiv preprint astro-ph/9712310

\bibitem[{R.~J. Campello {et~al.}(2013)Campello, Moulavi, \& Sander}]{Campello2013}
Campello, R.~J., Moulavi, D., \& Sander, J. 2013, in Density-based clustering based on hierarchical density estimates, Vol. 7819 LNAI, \dodoi{10.1007/978-3-642-37456-2_14}

\bibitem[{T. Cantat-Gaudin {et~al.}(2018)Cantat-Gaudin, Jordi, Vallenari, Bragaglia, Balaguer-Núñez, Soubiran, Bossini, Moitinho, Castro-Ginard, Krone-Martins, Casamiquela, Sordo, \& Carrera}]{Cantat2018}
Cantat-Gaudin, T., Jordi, C., Vallenari, A., {et~al.} 2018, \bibinfo{title}{A Gaia DR2 view of the open cluster population in the Milky Way,} Astronomy and Astrophysics, 618, \dodoi{10.1051/0004-6361/201833476}

\bibitem[{J. Coronado {et~al.}(2022)Coronado, Fürnkranz, \& Rix}]{Coronado2022}
Coronado, J., Fürnkranz, V., \& Rix, H.-W. 2022, \bibinfo{title}{Pearls on a String: Numerous Stellar Clusters Strung Along the Same Orbit,} The Astrophysical Journal, 928, \dodoi{10.3847/1538-4357/ac545c}

\bibitem[{J. Coronado {et~al.}(2020)Coronado, Rix, Trick, El-Badry, Rybizki, \& Xiang}]{Coronado2020}
Coronado, J., Rix, H.~W., Trick, W.~H., {et~al.} 2020, \bibinfo{title}{From birth associations to field stars: Mapping the small-scale orbit distribution in the Galactic disc,} Monthly Notices of the Royal Astronomical Society, 495, \dodoi{10.1093/MNRAS/STAA1358}

\bibitem[{O.~L. Creevey {et~al.}(2023)Creevey, Sarro, Lobel, Pancino, Andrae, Smart, Clementini, Heiter, Korn, Fouesneau, Frémat, Angeli, Vallenari, Harrison, Thévenin, Reylé, Sordo, Garofalo, Brown, Eyer, Prusti, Bruijne, Arenou, Babusiaux, Biermann, Ducourant, Wevers, Wyrzykowski, Yoldas, Yvard, Zhao, Zorec, Zucker, \& Zwitter}]{Creevey2023}
Creevey, O.~L., Sarro, L.~M., Lobel, A., {et~al.} 2023, \bibinfo{title}{Gaia Data Release 3: A golden sample of astrophysical parameters,} Astronomy and Astrophysics, 674, \dodoi{10.1051/0004-6361/202243800}

\bibitem[{D.~F. Crouse(2016)Crouse}]{linear_sum_assignment}
Crouse, D.~F. 2016, \bibinfo{title}{On implementing 2D rectangular assignment algorithms,} IEEE Transactions on Aerospace and Electronic Systems, 52, 1679, \dodoi{10.1109/TAES.2016.140952}

\bibitem[{W. Dias {et~al.}(2014)Dias, Monteiro, Caetano, L{\'e}pine, Assafin, \& Oliveira}]{dias2014proper}
Dias, W., Monteiro, H., Caetano, T., {et~al.} 2014, \bibinfo{title}{Proper motions of the optically visible open clusters based on the UCAC4 catalog,} Astronomy \& Astrophysics, 564, A79, \dodoi{10.1051/0004-6361/201323226}

\bibitem[{S.~W. Digel {et~al.}(1996)Digel, Lyder, Philbrick, Puche, \& Thaddeus}]{Digel1996}
Digel, S.~W., Lyder, D.~A., Philbrick, A.~J., Puche, D., \& Thaddeus, P. 1996, \bibinfo{title}{A Large-Scale CO Survey toward W3, W4, and W5,} The Astrophysical Journal, 458, \dodoi{10.1086/176839}

\bibitem[{ DR3(2022)DR3}]{DR3data}
DR3. 2022, \bibinfo{title}{Gaia DR3,} European Space Agency, \dodoi{10.5270/esa-qa4lep3}

\bibitem[{ EDR3(2020)EDR3}]{EDR3data}
EDR3. 2020, \bibinfo{title}{Gaia EDR3,} European Space Agency, \dodoi{10.5270/esa-1ugzkg7}

\bibitem[{M. Ester {et~al.}(1996)Ester, Kriegel, Sander, Xu, {et~al.}}]{ester1996density}
Ester, M., Kriegel, H.-P., Sander, J., Xu, X., {et~al.} 1996, in Proceedings of the Second International Conference on Knowledge Discovery and Data Mining (KDD), 226--231

\bibitem[{C. Fabricius {et~al.}(2021)Fabricius, Luri, Arenou, Babusiaux, Helmi, Muraveva, Reylé, Spoto, Vallenari, Antoja, Balbinot, Barache, Bauchet, Bragaglia, Busonero, Cantat-Gaudin, Carrasco, Diakité, Fabrizio, Figueras, Garcia-Gutierrez, Garofalo, Jordi, Kervella, Khanna, Leclerc, Licata, Lambert, Marrese, Masip, Ramos, Robichon, Robin, Romero-Gómez, Rubele, \& Weiler}]{Fabricius2021}
Fabricius, C., Luri, X., Arenou, F., {et~al.} 2021, \bibinfo{title}{Gaia Early Data Release 3: Catalogue validation,} Astronomy and Astrophysics, 649, \dodoi{10.1051/0004-6361/202039834}

\bibitem[{F.~A. Ferreira {et~al.}(2020)Ferreira, Corradi, Maia, Angelo, \& Santos}]{Ferreira2020}
Ferreira, F.~A., Corradi, W.~J., Maia, F.~F., Angelo, M.~S., \& Santos, J.~F. 2020, \bibinfo{title}{Discovery and astrophysical properties of Galactic open clusters in dense stellar fields using Gaia DR2,} Monthly Notices of the Royal Astronomical Society, 496, \dodoi{10.1093/MNRAS/STAA1684}

\bibitem[{M. Fouesneau {et~al.}(2023)Fouesneau, Fr{\'e}mat, Andrae, Korn, Soubiran, Kordopatis, Vallenari, Heiter, Creevey, Sarro, {et~al.}}]{fouesneau2023gaia}
Fouesneau, M., Fr{\'e}mat, Y., Andrae, R., {et~al.} 2023, \bibinfo{title}{Gaia Data Release 3-Apsis. II. Stellar parameters,} Astronomy \& Astrophysics, 674, A28

\bibitem[{ {Gaia Collaboration} {et~al.}(2016){Gaia Collaboration}, {Prusti}, {de Bruijne}, {Brown}, {Vallenari}, {Babusiaux}, {Bailer-Jones}, {Bastian}, {Biermann}, {Evans}, {Eyer}, {Jansen}, {Jordi}, {Klioner}, {Lammers}, {Lindegren}, {Luri}, {Mignard}, {Milligan}, {Panem}, {Poinsignon}, {Pourbaix}, {Randich}, {Sarri}, {Sartoretti}, {Siddiqui}, {Soubiran}, {Valette}, {van Leeuwen}, {Walton}, {Aerts}, {Arenou}, {Cropper}, {Drimmel}, {H{\o}g}, {Katz}, {Lattanzi}, {O'Mullane}, {Grebel}, {Holland}, {Huc}, {Passot}, {Bramante}, {Cacciari}, {Casta{\~n}eda}, {Chaoul}, {Cheek}, {De Angeli}, {Fabricius}, {Guerra}, {Hern{\'a}ndez}, {Jean-Antoine-Piccolo}, {Masana}, {Messineo}, {Mowlavi}, {Nienartowicz}, {Ord{\'o}{\~n}ez-Blanco}, {Panuzzo}, {Portell}, {Richards}, {Riello}, {Seabroke}, {Tanga}, {Th{\'e}venin}, {Torra}, {Els}, {Gracia-Abril}, {Comoretto}, {Garcia-Reinaldos}, {Lock}, {Mercier}, {Altmann}, {Andrae}, {Astraatmadja}, {Bellas-Velidis}, {Benson}, {Berthier}, {Blomme}, {Busso}, {Carry}, {Cellino},
  {Clementini}, {Cowell}, {Creevey}, {Cuypers}, {Davidson}, {De Ridder}, {de Torres}, {Delchambre}, {Dell'Oro}, {Ducourant}, {Fr{\'e}mat}, {Garc{\'\i}a-Torres}, {Gosset}, {Halbwachs}, {Hambly}, {Harrison}, {Hauser}, {Hestroffer}, {Hodgkin}, {Huckle}, {Hutton}, {Jasniewicz}, {Jordan}, {Kontizas}, {Korn}, {Lanzafame}, {Manteiga}, {Moitinho}, {Muinonen}, {Osinde}, {Pancino}, {Pauwels}, {Petit}, {Recio-Blanco}, {Robin}, {Sarro}, {Siopis}, {Smith}, {Smith}, {Sozzetti}, {Thuillot}, {van Reeven}, {Viala}, {Abbas}, {Abreu Aramburu}, {Accart}, {Aguado}, {Allan}, {Allasia}, {Altavilla}, {{\'A}lvarez}, {Alves}, {Anderson}, {Andrei}, {Anglada Varela}, {Antiche}, {Antoja}, {Ant{\'o}n}, {Arcay}, {Atzei}, {Ayache}, {Bach}, {Baker}, {Balaguer-N{\'u}{\~n}ez}, {Barache}, {Barata}, {Barbier}, {Barblan}, {Baroni}, {Barrado y Navascu{\'e}s}, {Barros}, {Barstow}, {Becciani}, {Bellazzini}, {Bellei}, {Bello Garc{\'\i}a}, {Belokurov}, {Bendjoya}, {Berihuete}, {Bianchi}, {Bienaym{\'e}}, {Billebaud}, {Blagorodnova}, {Blanco-Cuaresma},
  {Boch}, {Bombrun}, {Borrachero}, {Bouquillon}, {Bourda}, {Bouy}, {Bragaglia}, {Breddels}, {Brouillet}, {Br{\"u}semeister}, {Bucciarelli}, {Budnik}, {Burgess}, {Burgon}, {Burlacu}, {Busonero}, {Buzzi}, {Caffau}, {Cambras}, {Campbell}, {Cancelliere}, {Cantat-Gaudin}, {Carlucci}, {Carrasco}, {Castellani}, {Charlot}, {Charnas}, {Charvet}, {Chassat}, {Chiavassa}, {Clotet}, {Cocozza}, {Collins}, {Collins}, \& {Costigan}}]{gaia_2016b}
{Gaia Collaboration}, {Prusti}, T., {de Bruijne}, J.~H.~J., {et~al.} 2016, \bibinfo{title}{{The Gaia mission},} \aap, 595, A1, \dodoi{10.1051/0004-6361/201629272}

\bibitem[{ {Gaia Collaboration} {et~al.}(2023){Gaia Collaboration}, {Vallenari}, {Brown}, {Prusti}, {de Bruijne}, {Arenou}, {Babusiaux}, {Biermann}, {Creevey}, {Ducourant}, {Evans}, {Eyer}, {Guerra}, {Hutton}, {Jordi}, {Klioner}, {Lammers}, {Lindegren}, {Luri}, {Mignard}, {Panem}, {Pourbaix}, {Randich}, {Sartoretti}, {Soubiran}, {Tanga}, {Walton}, {Bailer-Jones}, {Bastian}, {Drimmel}, {Jansen}, {Katz}, {Lattanzi}, {van Leeuwen}, {Bakker}, {Cacciari}, {Casta{\~n}eda}, {De Angeli}, {Fabricius}, {Fouesneau}, {Fr{\'e}mat}, {Galluccio}, {Guerrier}, {Heiter}, {Masana}, {Messineo}, {Mowlavi}, {Nicolas}, {Nienartowicz}, {Pailler}, {Panuzzo}, {Riclet}, {Roux}, {Seabroke}, {Sordo}, {Th{\'e}venin}, {Gracia-Abril}, {Portell}, {Teyssier}, {Altmann}, {Andrae}, {Audard}, {Bellas-Velidis}, {Benson}, {Berthier}, {Blomme}, {Burgess}, {Busonero}, {Busso}, {C{\'a}novas}, {Carry}, {Cellino}, {Cheek}, {Clementini}, {Damerdji}, {Davidson}, {de Teodoro}, {Nu{\~n}ez Campos}, {Delchambre}, {Dell'Oro}, {Esquej},
  {Fern{\'a}ndez-Hern{\'a}ndez}, {Fraile}, {Garabato}, {Garc{\'\i}a-Lario}, {Gosset}, {Haigron}, {Halbwachs}, {Hambly}, {Harrison}, {Hern{\'a}ndez}, {Hestroffer}, {Hodgkin}, {Holl}, {Jan{\ss}en}, {Jevardat de Fombelle}, {Jordan}, {Krone-Martins}, {Lanzafame}, {L{\"o}ffler}, {Marchal}, {Marrese}, {Moitinho}, {Muinonen}, {Osborne}, {Pancino}, {Pauwels}, {Recio-Blanco}, {Reyl{\'e}}, {Riello}, {Rimoldini}, {Roegiers}, {Rybizki}, {Sarro}, {Siopis}, {Smith}, {Sozzetti}, {Utrilla}, {van Leeuwen}, {Abbas}, {{\'A}brah{\'a}m}, {Abreu Aramburu}, {Aerts}, {Aguado}, {Ajaj}, {Aldea-Montero}, {Altavilla}, {{\'A}lvarez}, {Alves}, {Anders}, {Anderson}, {Anglada Varela}, {Antoja}, {Baines}, {Baker}, {Balaguer-N{\'u}{\~n}ez}, {Balbinot}, {Balog}, {Barache}, {Barbato}, {Barros}, {Barstow}, {Bartolom{\'e}}, {Bassilana}, {Bauchet}, {Becciani}, {Bellazzini}, {Berihuete}, {Bernet}, {Bertone}, {Bianchi}, {Binnenfeld}, {Blanco-Cuaresma}, {Blazere}, {Boch}, {Bombrun}, {Bossini}, {Bouquillon}, {Bragaglia}, {Bramante}, {Breedt},
  {Bressan}, {Brouillet}, {Brugaletta}, {Bucciarelli}, {Burlacu}, {Butkevich}, {Buzzi}, {Caffau}, {Cancelliere}, {Cantat-Gaudin}, {Carballo}, {Carlucci}, {Carnerero}, {Carrasco}, {Casamiquela}, {Castellani}, {Castro-Ginard}, {Chaoul}, {Charlot}, {Chemin}, {Chiaramida}, {Chiavassa}, {Chornay}, {Comoretto}, {Contursi}, {Cooper}, {Cornez}, {Cowell}, {Crifo}, {Cropper}, {Crosta}, {Crowley}, {Dafonte}, {Dapergolas}, {David}, {David}, {de Laverny}, {De Luise}, \& {De March}}]{gaia_2023j}
{Gaia Collaboration}, {Vallenari}, A., {Brown}, A.~G.~A., {et~al.} 2023, \bibinfo{title}{{Gaia Data Release 3. Summary of the content and survey properties},} \aap, 674, A1, \dodoi{10.1051/0004-6361/202243940}

\bibitem[{ {Gaia Collaboration (Drimmel, R., et al.)}(2023a){Gaia Collaboration (Drimmel, R., et al.)}}]{drimmel2023gaia}
{Gaia Collaboration (Drimmel, R., et al.)}. 2023a, \bibinfo{title}{Gaia Data Release 3-Mapping the asymmetric disc of the Milky Way,} Astronomy \& astrophysics, 674, A37, \dodoi{10.1051/0004-6361/202243797}

\bibitem[{ {Gaia Collaboration (Recio-Blanco, A., et al.)}(2023b){Gaia Collaboration (Recio-Blanco, A., et al.)}}]{recio2023gaia}
{Gaia Collaboration (Recio-Blanco, A., et al.)}. 2023b, \bibinfo{title}{Gaia Data Release 3-Chemical cartography of the Milky Way,} Astronomy \& Astrophysics, 674, A38, \dodoi{10.1051/0004-6361/202243511}

\bibitem[{ {Gaia Collaboration (Schultheis, M. et al.)}(2023c){Gaia Collaboration (Schultheis, M. et al.)}}]{schultheis2023gaia}
{Gaia Collaboration (Schultheis, M. et al.)}. 2023c, \bibinfo{title}{Gaia Data Release 3-Exploring and mapping the diffuse interstellar band at 862 nm,} Astronomy \& astrophysics, 674, A40, \dodoi{10.1051/0004-6361/202243283}

\bibitem[{ {GNU Project}(2025){GNU Project}}]{gnu_scientific_library}
{GNU Project}. 2025, GNU Scientific Library -- Reference Manual: Weighted Samples, gnu.org.
\newblock \url{https://www.gnu.org/software/gsl/doc/html/statistics.html#weighted-samples}

\bibitem[{Z.~H. He {et~al.}(2021)He, Xu, Hao, Wu, \& Li}]{He2021}
He, Z.~H., Xu, Y., Hao, C.~J., Wu, Z.~Y., \& Li, J.~J. 2021, \bibinfo{title}{A catalogue of 74 new open clusters found in Gaia Data-Release 2,} Research in Astronomy and Astrophysics, 21, \dodoi{10.1088/1674-4527/21/4/93}

\bibitem[{M. {Heyer} \& T.~M. {Dame}(2015){Heyer} \& {Dame}}]{heyer2015molecular}
{Heyer}, M., \& {Dame}, T.~M. 2015, \bibinfo{title}{{Molecular Clouds in the Milky Way},} \araa, 53, 583, \dodoi{10.1146/annurev-astro-082214-122324}

\bibitem[{E.~L. Hunt \& S. Reffert(2023)Hunt \& Reffert}]{Hunt2023}
Hunt, E.~L., \& Reffert, S. 2023, \bibinfo{title}{Improving the open cluster census,} Astronomy \& Astrophysics, 673, \dodoi{10.1051/0004-6361/202346285}

\bibitem[{J.~D. Hunter(2007)Hunter}]{Hunter:2007}
Hunter, J.~D. 2007, \bibinfo{title}{Matplotlib: A 2D graphics environment,} Computing in Science \& Engineering, 9, 90, \dodoi{10.1109/MCSE.2007.55}

\bibitem[{ {Inkscape Project}(2020){Inkscape Project}}]{Inkscape}
{Inkscape Project}. 2020, \bibinfo{title}{Inkscape,}, 0.92.5 \url{https://inkscape.org}

\bibitem[{Z. Jun-liang {et~al.}(1982)Jun-liang, Kai-ping, Zong-hai, \& Ming-guan}]{jun1982discussion}
Jun-liang, Z., Kai-ping, T., Zong-hai, X., \& Ming-guan, Y. 1982, \bibinfo{title}{Discussion on the maximum likelihood method for determination of membership in open clusters,} Chinese Astronomy and Astrophysics, 6, 293, \dodoi{10.1016/0275-1062(82)90004-2}

\bibitem[{R. Kerr {et~al.}(2023)Kerr, Kraus, \& Rizzuto}]{Kerr2023}
Kerr, R., Kraus, A.~L., \& Rizzuto, A.~C. 2023, \bibinfo{title}{SPYGLASS. IV. New Stellar Survey of Recent Star Formation within 1 kpc,} The Astrophysical Journal, 954, \dodoi{10.3847/1538-4357/ace5b3}

\bibitem[{I. King(1962)King}]{King1962}
King, I. 1962, \bibinfo{title}{The structure of star clusters. I. an empirical density law,} The Astronomical Journal, 67, \dodoi{10.1086/108756}

\bibitem[{D.~M. Krolikowski {et~al.}(2021)Krolikowski, Kraus, \& Rizzuto}]{Krolikowski2021}
Krolikowski, D.~M., Kraus, A.~L., \& Rizzuto, A.~C. 2021, \bibinfo{title}{Gaia EDR3 Reveals the Substructure and Complicated Star Formation History of the Greater Taurus-Auriga Star-forming Complex,} The Astronomical Journal, 162, \dodoi{10.3847/1538-3881/ac0632}

\bibitem[{C.~J. Lada \& E.~A. Lada(2003)Lada \& Lada}]{lada2003embedded}
Lada, C.~J., \& Lada, E.~A. 2003, \bibinfo{title}{Embedded clusters in molecular clouds,} Annual Review of Astronomy and Astrophysics, 41, 57

\bibitem[{L. Liu \& X. Pang(2019)Liu \& Pang}]{Liu2019}
Liu, L., \& Pang, X. 2019, \bibinfo{title}{A Catalog of Newly Identified Star Clusters in Gaia DR2,} The Astrophysical Journal Supplement Series, 245, \dodoi{10.3847/1538-4365/ab530a}

\bibitem[{G. Marton {et~al.}(2023)Marton, {\'A}brah{\'a}m, Rimoldini, Audard, Kun, Nagy, K{\'o}sp{\'a}l, Szabados, Holl, Gavras, {et~al.}}]{marton2023gaia}
Marton, G., {\'A}brah{\'a}m, P., Rimoldini, L., {et~al.} 2023, \bibinfo{title}{Gaia Data Release 3-Validating the classification of variable young stellar object candidates,} Astronomy \& Astrophysics, 674, A21

\bibitem[{L. McInnes \& J. Healy(2017)McInnes \& Healy}]{mcinnes2017accelerated}
McInnes, L., \& Healy, J. 2017, in Data Mining Workshops (ICDMW), 2017 IEEE International Conference on, IEEE, 33--42

\bibitem[{L. McInnes {et~al.}(2017)McInnes, Healy, \& Astels}]{McInnes2017}
McInnes, L., Healy, J., \& Astels, S. 2017, \bibinfo{title}{hdbscan: Hierarchical density based clustering,} The Journal of Open Source Software, 2, \dodoi{10.21105/joss.00205}

\bibitem[{C.~F. {McKee} \& E.~C. {Ostriker}(2007){McKee} \& {Ostriker}}]{mckee2007theory}
{McKee}, C.~F., \& {Ostriker}, E.~C. 2007, \bibinfo{title}{{Theory of Star Formation},} \araa, 45, 565, \dodoi{10.1146/annurev.astro.45.051806.110602}

\bibitem[{K.~N. {Mead} \& M.~L. {Kutner}(1988){Mead} \& {Kutner}}]{mead1988molecular}
{Mead}, K.~N., \& {Kutner}, M.~L. 1988, \bibinfo{title}{{Molecular Clouds in the Outer Galaxy. III. CO Studies of Individual Clouds},} \apj, 330, 399, \dodoi{10.1086/166479}

\bibitem[{V. Patel {et~al.}(2025)Patel, Hora, Ashby, \& Vig}]{patel2025}
Patel, V., Hora, J., Ashby, M., \& Vig, S. 2025, \bibinfo{title}{HDBSCAN-MC and GOLDFIST Simulation,}, 1.0.0 Zenodo, \dodoi{10.5281/zenodo.15511563}

\bibitem[{G.~I. Perren {et~al.}(2020)Perren, Giorgi, Moitinho, Carraro, Pera, \& Vázquez}]{Perren2020}
Perren, G.~I., Giorgi, E.~E., Moitinho, A., {et~al.} 2020, \bibinfo{title}{Sixteen overlooked open clusters in the fourth Galactic quadrant,} Astronomy \& Astrophysics, 637, \dodoi{10.1051/0004-6361/201937141}

\bibitem[{G.~I. Perren {et~al.}(2015)Perren, Vázquez, \& Piatti}]{Perren2015}
Perren, G.~I., Vázquez, R.~A., \& Piatti, A.~E. 2015, \bibinfo{title}{ASteCA: Automated Stellar Cluster Analysis,} Astronomy and Astrophysics, 576, \dodoi{10.1051/0004-6361/201424946}

\bibitem[{M.~A. Perryman {et~al.}(2001)Perryman, Boer, Gilmore, Høg, Lattanzi, Lindegren, Luri, Mignard, Pace, \& Zeeuw}]{Perryman2001}
Perryman, M.~A., Boer, K. S.~D., Gilmore, G., {et~al.} 2001, \bibinfo{title}{GAIA: Composition, formation and evolution of the Galaxy,} Astronomy and Astrophysics, 369, \dodoi{10.1051/0004-6361:20010085}

\bibitem[{M.~F. Qin {et~al.}(2023)Qin, Zhang, Liu, Song, Hu, Wang, Ma, \& Lü}]{Qin2023}
Qin, M.~F., Zhang, Y., Liu, J., {et~al.} 2023, \bibinfo{title}{Cluster aggregates surrounding Pismis 5 in the Vela molecular ridge,} Astronomy \& Astrophysics, 675, A67, \dodoi{10.1051/0004-6361/202244737}

\bibitem[{L. Rimoldini {et~al.}(2023)Rimoldini, Holl, Gavras, Audard, Ridder, Mowlavi, Nienartowicz, Fombelle, Lecoeur-Taïbi, Karbevska, Evans, Ábrahám, Carnerero, Clementini, Distefano, Garofalo, García-Lario, Gomel, Klioner, Kruszyńska, Lanzafame, Lebzelter, Marton, Mazeh, Molinaro, Panahi, Raiteri, Ripepi, Szabados, Teyssier, Trabucchi, Łukasz Wyrzykowski, Zucker, \& Eyer}]{Rimoldini2023}
Rimoldini, L., Holl, B., Gavras, P., {et~al.} 2023, \bibinfo{title}{Gaia Data Release 3: All-sky classification of 12.4 million variable sources into 25 classes,} Astronomy and Astrophysics, 674, \dodoi{10.1051/0004-6361/202245591}

\bibitem[{A.~L. {Rudolph} {et~al.}(1997){Rudolph}, {Simpson}, {Haas}, {Erickson}, \& {Fich}}]{rudolph1997far}
{Rudolph}, A.~L., {Simpson}, J.~P., {Haas}, M.~R., {Erickson}, E.~F., \& {Fich}, M. 1997, \bibinfo{title}{{Far-Infrared Abundance Measurements in the Outer Galaxy},} \apj, 489, 94, \dodoi{10.1086/304758}

\bibitem[{D.~W. Scott(2015)Scott}]{scott2015multivariate}
Scott, D.~W. 2015, Multivariate density estimation: theory, practice, and visualization (John Wiley \& Sons), \dodoi{10.1002/9781118575574}

\bibitem[{V. Sherina(2014)Sherina}]{sherina2014generalized}
Sherina, V. 2014, \bibinfo{title}{Generalized Weibull and Inverse Weibull Distributions with Applications,}

\bibitem[{G. Sim {et~al.}(2019)Sim, Lee, Ann, \& Kim}]{Sim2019}
Sim, G., Lee, S.~H., Ann, H.~B., \& Kim, S. 2019, \bibinfo{title}{207 new open star clusters within 1 kpc from gaia data release 2,} Journal of the Korean Astronomical Society, 52, \dodoi{10.5303/JKAS.2019.52.5.145}

\bibitem[{M.~F. {Skrutskie} {et~al.}(2006){Skrutskie}, {Cutri}, {Stiening}, {Weinberg}, {Schneider}, {Carpenter}, {Beichman}, {Capps}, {Chester}, {Elias}, {Huchra}, {Liebert}, {Lonsdale}, {Monet}, {Price}, {Seitzer}, {Jarrett}, {Kirkpatrick}, {Gizis}, {Howard}, {Evans}, {Fowler}, {Fullmer}, {Hurt}, {Light}, {Kopan}, {Marsh}, {McCallon}, {Tam}, {Van Dyk}, \& {Wheelock}}]{2mass}
{Skrutskie}, M.~F., {Cutri}, R.~M., {Stiening}, R., {et~al.} 2006, \bibinfo{title}{{The Two Micron All Sky Survey (2MASS)},} \aj, 131, 1163, \dodoi{10.1086/498708}

\bibitem[{R. Storn \& K. Price(1995)Storn \& Price}]{storn1995differential}
Storn, R., \& Price, K. 1995, \bibinfo{title}{Differential evolution-a simple and efficient adaptive scheme for global optimization over continuous spaces,} International computer science institute

\bibitem[{M.~B. Taylor(2005)Taylor}]{taylor2005topcat}
Taylor, M.~B. 2005, in Astronomical data analysis software and systems XIV, Vol. 347, 29

\bibitem[{R.~J. {Trumpler}(1930){Trumpler}}]{trumpler1979preliminary}
{Trumpler}, R.~J. 1930, \bibinfo{title}{{Preliminary results on the distances, dimensions and space distribution of open star clusters},} Lick Observatory Bulletin, 420, 154, \dodoi{10.5479/ADS/bib/1930LicOB.14.154T}

\bibitem[{P. Virtanen {et~al.}(2020)Virtanen, Gommers, Oliphant, Haberland, Reddy, Cournapeau, Burovski, Peterson, Weckesser, Bright, {van der Walt}, Brett, Wilson, Millman, Mayorov, Nelson, Jones, Kern, Larson, Carey, Polat, Feng, Moore, {VanderPlas}, Laxalde, Perktold, Cimrman, Henriksen, Quintero, Harris, Archibald, Ribeiro, Pedregosa, {van Mulbregt}, \& {SciPy 1.0 Contributors}}]{2020SciPy-NMeth}
Virtanen, P., Gommers, R., Oliphant, T.~E., {et~al.} 2020, \bibinfo{title}{{{SciPy} 1.0: Fundamental Algorithms for Scientific Computing in Python},} Nature Methods, 17, 261, \dodoi{10.1038/s41592-019-0686-2}

\bibitem[{S.~N. Vogel {et~al.}(1988)Vogel, Kulkarni, \& Scoville}]{Vogel1988}
Vogel, S.~N., Kulkarni, S.~R., \& Scoville, N.~Z. 1988, \bibinfo{title}{Star formation in giant molecular associations synchronized by a spiral density wave,} Nature, 334, \dodoi{10.1038/334402a0}

\bibitem[{M. Waskom {et~al.}(2017)Waskom, Botvinnik, O'Kane, Hobson, Lukauskas, Gemperline, Augspurger, Halchenko, Cole, Warmenhoven, de~Ruiter, Pye, Hoyer, Vanderplas, Villalba, Kunter, Quintero, Bachant, Martin, Meyer, Miles, Ram, Yarkoni, Williams, Evans, Fitzgerald, Brian, Fonnesbeck, Lee, \& Qalieh}]{seaborn}
Waskom, M., Botvinnik, O., O'Kane, D., {et~al.} 2017, \bibinfo{title}{mwaskom/seaborn: v0.8.1 (September 2017),}, v0.8.1 Zenodo, \dodoi{10.5281/zenodo.883859}

\bibitem[{M.~W. {Werner} {et~al.}(2004){Werner}, {Roellig}, {Low}, {Rieke}, {Rieke}, {Hoffmann}, {Young}, {Houck}, {Brandl}, {Fazio}, {Hora}, {Gehrz}, {Helou}, {Soifer}, {Stauffer}, {Keene}, {Eisenhardt}, {Gallagher}, {Gautier}, {Irace}, {Lawrence}, {Simmons}, {Van Cleve}, {Jura}, {Wright}, \& {Cruikshank}}]{spitzer}
{Werner}, M.~W., {Roellig}, T.~L., {Low}, F.~J., {et~al.} 2004, \bibinfo{title}{{The Spitzer Space Telescope Mission},} \apjs, 154, 1, \dodoi{10.1086/422992}

\bibitem[{E. Winston {et~al.}(2020)Winston, Hora, \& Tolls}]{Winston2020}
Winston, E., Hora, J.~L., \& Tolls, V. 2020, A Census of Star Formation in the Outer Galaxy. II. The GLIMPSE360 Field, Vol. 160, \dodoi{10.3847/1538-3881/ab99c8}

\bibitem[{E.~L. {Wright} {et~al.}(2010){Wright}, {Eisenhardt}, {Mainzer}, {Ressler}, {Cutri}, {Jarrett}, {Kirkpatrick}, {Padgett}, {McMillan}, {Skrutskie}, {Stanford}, {Cohen}, {Walker}, {Mather}, {Leisawitz}, {Gautier}, {McLean}, {Benford}, {Lonsdale}, {Blain}, {Mendez}, {Irace}, {Duval}, {Liu}, {Royer}, {Heinrichsen}, {Howard}, {Shannon}, {Kendall}, {Walsh}, {Larsen}, {Cardon}, {Schick}, {Schwalm}, {Abid}, {Fabinsky}, {Naes}, \& {Tsai}}]{wise}
{Wright}, E.~L., {Eisenhardt}, P. R.~M., {Mainzer}, A.~K., {et~al.} 2010, \bibinfo{title}{{The Wide-field Infrared Survey Explorer (WISE): Mission Description and Initial On-orbit Performance},} \aj, 140, 1868, \dodoi{10.1088/0004-6256/140/6/1868}

\bibitem[{M. Zhang(2023)Zhang}]{Zhang2023}
Zhang, M. 2023, \bibinfo{title}{Distances to Nearby Molecular Clouds Traced by Young Stars,} The Astrophysical Journal Supplement Series, 265, 59, \dodoi{10.3847/1538-4365/acc1e8}

\end{thebibliography}
\bibliographystyle{aasjournal}

\end{document}